\documentclass[
	aps, prd, reprint,
	10pt, notitlepage, a4paper,
        floats, floatfix,
	amsmath, amssymb, amsfonts, eqsecnum,
	superscriptaddress,
	showpacs, showkeys,
	nofootinbib,
 	longbibliography,
]{revtex4-1}

\usepackage{graphicx}
\usepackage[usenames,dvipsnames]{xcolor}
\usepackage{tikz}
\usepackage{xspace} 
\usepackage{dcolumn}
\usepackage{bm}
\usepackage{multirow} 
\usepackage[utf8]{inputenc} 
\usepackage{latexsym}
\usepackage{float}

\xdefinecolor{mylinkcolor}{rgb}{0,0,0.5}
\usepackage[
	bookmarksnumbered, bookmarksopen, bookmarksopenlevel=2,
	breaklinks=true,
	colorlinks=true, filecolor=mylinkcolor, citecolor=mylinkcolor,
	linkcolor=mylinkcolor, urlcolor=mylinkcolor, menucolor=mylinkcolor,
]{hyperref}

\allowdisplaybreaks

\begin{document}

\hypersetup{ pdftitle={Canonical Hamiltonian for an extended test body in curved spacetime: To quadratic order in spin},
pdfauthor={Justin Vines, Daniela Kunst, Jan Steinhoff, Tanja Hinderer} }

\title{Canonical Hamiltonian for an extended test body in curved spacetime:\\To quadratic order in spin}

\author{Justin Vines}

\email{justin.vines@aei.mpg.de}

\affiliation{Max-Planck-Institute for Gravitational Physics (Albert-Einstein-Institute),
Am M{\"u}hlenberg 1, 14476 Potsdam-Golm, Germany, EU}

\author{Daniela Kunst}

\email{daniela.kunst@zarm.uni-bremen.de}

\affiliation{ZARM, University of Bremen, Am Fallturm, 28359 Bremen, Germany, EU}

\author{Jan Steinhoff}

\email{jan.steinhoff@aei.mpg.de}

\homepage{http://jan-steinhoff.de/physics/}

\affiliation{Max-Planck-Institute for Gravitational Physics (Albert-Einstein-Institute),
Am M{\"u}hlenberg 1, 14476 Potsdam-Golm, Germany, EU}

\affiliation{Centro Multidisciplinar de Astrof{\'i}sica --- CENTRA, Departamento
de F{\'i}sica, Instituto Superior T{\'e}cnico --- IST, Universidade de
Lisboa --- ULisboa, Avenida Rovisco Pais 1, 1049-001 Lisboa, Portugal, EU}

\author{Tanja Hinderer}

\email{tanjah@umd.edu}

\affiliation{Maryland Center for Fundamental Physics \& Joint Space-Science Institute,
Department of Physics, University of Maryland, College Park, MD 20742,
USA}

\begin{abstract}
We derive a Hamiltonian for an extended spinning test body in a curved background spacetime, to quadratic order in the spin, in terms of three-dimensional position, momentum, and spin variables having canonical Poisson brackets.  This requires a careful analysis of how changes of the spin supplementary condition are related to shifts of the body's representative worldline and transformations of the body's multipole moments, and we employ bitensor calculus for a precise framing of this analysis.  We apply the result to the case of the Kerr spacetime and thereby compute an explicit canonical Hamiltonian for the test-body limit of the spinning two-body problem in general relativity, valid for generic orbits and spin orientations, to quadratic order in the test spin.  This fully relativistic Hamiltonian is then expanded in post-Newtonian orders and in powers of the Kerr spin parameter, allowing comparisons with the test-mass limits of available post-Newtonian results.  Both the fully relativistic Hamiltonian and the results of its expansion can inform the construction of waveform models, especially effective-one-body models, for the analysis of gravitational waves from compact binaries.
\end{abstract}

\date{\today}

\maketitle

\section{Introduction}

The advent of gravitational wave astronomy---which has commenced with the first detection of a binary black hole merger \cite{detectionpaper}---promises to shed light on many profound questions in astrophysics and gravitational physics.
The first such questions within the reach of gravitational wave observatories will concern the nature of gravity in the strong-field regime and the properties of black holes and neutron stars, as inspiraling and coalescing binary systems of such compact objects should be frequent sources,
 both for the advanced generation of ground-based gravitational wave detectors~\cite{Shoemaker2009, Acernese-etal:2006,Kuroda:2010} and for future space-based detectors~\cite{eLISA}, as well as for pulsar timing arrays \cite{IPTA}.  Understanding in great detail the dynamics of such two-body systems, expected to be governed by general relativity, is thus a cornerstone objective of gravitational wave physics.

A sufficiently accurate and general solution to the relativistic two-body problem will require a synergy of results from both numerical and analytic computations.  On the analytic side, two complementary approximation schemes are available: the post-Newtonian (PN) approximation expands about the Newtonian limit but is valid for arbitrary mass ratios~\cite{Blanchet:2006, Futamase:2007zz}, while the extreme-mass-ratio (EMR) approximation expands about the test-mass limit but is valid in the strong-field, relativistic regime.  A synergistic approach is the effective-one-body (EOB) formalism~\cite{Buonanno99,Buonanno00}, which incorporates information from the PN limit, the EMR limit, and numerical relativity in an attempt to provide an accurate effective description of two-body systems throughout the parameter space.

In the EMR approximation, the zeroth-order solution is given by a (point) test mass moving along a geodesic of a background black hole spacetime---the Schwarzschild spacetime of a nonspinning black hole or the Kerr spacetime of a spinning black hole.  Corrections to this solution can proceed in two (intermingled) directions:  Firstly, one can compute the perturbation to the gravitational field produced by the small body, its self-field, and the resultant influence on its motion.  This is the goal of the ``self-force'' paradigm, as reviewed e.g.\ by Refs.~\cite{Poisson:Pound:Vega:2011,Barack:2014,Pound_review}.  Secondly, one can compute ``finite-size effects'' on the small body's motion, due to its spin and to intrinsic and tidally induced deformations.  Such finite-size effects in the EMR limit (neglecting the self-field) are the focus of this paper.

The equations of motion of a spinning (pole-dipole) test body in curved spacetime were first derived by Mathisson~\cite{Mathisson:1937,Mathisson:2010} and Papapetrou~\cite{Papapetrou:1951pa} and were later extended to include the effects of higher multipoles by Dixon \cite{Dixon:1979}; see \cite{Dixon:2015vxa} for a review.
The resultant dynamics of a spinning test body (to pole-dipole order) serves as the basis of the spinning EOB models of Refs.~\cite{Barausse:2009xi,Barausse:2011ys,taracchini2012,taracchini2014,pan2014}, which employ the canonical Hamiltonian for a pole-dipole particle derived in Ref.~\cite{Barausse:Racine:Buonanno:2009}.  The conservative dynamics of these EOB models is defined by the Hamiltonian of an effective spinning test particle in an effective spacetime which is a deformation of the Kerr spacetime, in the same way that the original EOB model \cite{Buonanno99,Buonanno00} was based on a nonspinning test particle in a deformation of the Schwarzschild spacetime.  In both cases, the deformations encode finite-mass-ratio effects determined from PN calculations and vanish as the mass ratio goes to zero, so that exact (strong-field) results are recovered in the test-particle limit. We should note that other successful EOB models, including the original spinning EOB models \cite{Damour1SEOB,Damour:2008qf}, and more recently, e.g., Refs.~\cite{Damour:2014sva,Nagar:2011fx,Balmelli:2015zsa}, do not incorporate the exact spinning test-particle limit (but do include the geodesic test-mass limit).

This paper extends the work of Ref.~\cite{Barausse:Racine:Buonanno:2009}, which was valid to linear order in the spin, or to dipolar order in the body's multipole expansion, to derive a canonical Hamiltonian for an extended test body in a curved background which is valid to quadrupolar order.  We treat explicitly here spin-induced quadrupoles and all other spin-squared effects.  Our methods also provide significant simplifications of some of the dipole-order calculations of Ref.~\cite{Barausse:Racine:Buonanno:2009}, as we employ crucial insights from Ref.~\cite{Levi:Steinhoff:2015:1} on the handling of generic spin supplementary conditions (SSCs) [conditions which fix a representative center-of-mass worldline for the test body] at the level of the action.  We highlight how a change of the SSC, corresponding to a shift of the center-of-mass worldline, entails transformations of the body's multipole moments and corresponding modifications of the action and Hamiltonian.  We use bitensor calculus \cite{Poisson:Pound:Vega:2011, DeWitt:Brehme:1960,Synge} to provide a precise and manifestly covariant treatment of the worldline shift.  We finally show, extending the linear-in-spin result in Ref.~\cite{Barausse:Racine:Buonanno:2009}, how use of the Newton-Wigner SSC \cite{Pryce:1935, Pryce:1948, Newton:Wigner:1949} allows one to construct a Hamiltonian in terms of three-dimensional position, momentum, and spin variables with a canonical Poisson bracket structure.
Beside Ref.~\cite{Barausse:Racine:Buonanno:2009}, a canonical formalism for spinning test-particles in general relativity was obtained in \cite{Kunzle:1972} through a direct construction of the symplectic structure and in \cite{Hartung:Steinhoff:Schafer:2012} from an ADM canonical formulation.

The result for the canonical Hamiltonian can be summarized as follows.  In a spacetime with coordinates $x^\mu=(t,x^\mathrm{i})$ and an orthonormal frame (or tetrad) $e_a{}^\mu=(e_0{}^\mu,e_i{}^\mu)$, the (reduced) phase space for a spinning test body consists of the spatial coordinates $z^\mathrm{i}$ of its representative worldline $z^\mu$, their canonically conjugate momenta $P_{\,\mathrm{i}}$, and the frame spatial components $S_i=\frac{1}{2}\epsilon_{ijk}S_{jk}=\tfrac{1}{2}\epsilon_{ijk}e_j{}^\mu e_k{}^\nu S_{\mu\nu}$ of the spin tensor $S_{\mu\nu}$, given as functions of the time coordinate $t$ and obeying the canonical Poisson brackets (\ref{PBs}).  The Hamiltonian $H(z,P,S)$ is defined by
\begin{equation}\label{introH}
H=-P_t=N\sqrt{\mu^2+\gamma^\mathrm{ij}P_{\,\mathrm{i}}P_{\,\mathrm{j}}}-N^{\mathrm{i}}P_{\,\mathrm{i}},
\end{equation}
where $P_\mu=(P_t,P_{\,\mathrm{i}})$ is the 4D canonical momentum whose time component is the minus the Hamiltonian, where $N$, $N^{\mathrm{i}}$, and $\gamma^\mathrm{ij}$ are the lapse, shift, and inverse spatial metric as in (\ref{lapseshift}), and where the (canonical) \mbox{effective} dynamical mass $\mu(z,P,S)$ is given by (\ref{eq:msq}),
\begin{align}\label{intromu}
\mu^{2}&= -P_\mu P^\mu
\\*
&= m^{2}-P^a\omega_{a}{}^{bc}S_{bc}+\frac{1}{4}\omega_{a}{}^{bc}\omega^{ade}S_{bc}S_{de}
\nonumber\\*\nonumber
&\quad+\frac{1}{4}R_{abcd}S^{ab}S^{cd}-(C-1)E^{(P)}_{ab}s^as^b+\mathcal{O}(S^{3}).
\end{align}
Here, $\omega_{abc}=e_a{}^\mu(\nabla_\mu e_b{}^\nu)e_{c\nu}$ are the Ricci rotation coefficients, $C$ is a constant response coefficient measuring the test body's spin-induced quadrupolar deformation with $C=1$ for test black holes, $E^{(P)}_{ab}$ is the electric part of the Weyl/vacuum-Riemann tensor with respect to $P^a$ (\ref{EB}), $s^a$ is the Pauli-Lubanski spin vector (\ref{PLSV}), and the frame components $S_{0i}$ are determined by the solution (\ref{S0s}) of the Newton-Wigner SSC. The constant mass $m(S)$ is a function of the likewise constant spin length $S = \sqrt{s^a s_a}$ and encodes the moment of inertia \cite{Steinhoff:2014}.  

Our derivation of the Hamiltonian (and the covariant action principle which yields it) resolves some previous ambiguities concerning the adjustability of the coefficients of the curvature coupling terms in the last line of (\ref{intromu}), as we discuss in particular below (\ref{JmcM}) and below (\ref{masterM}).

By specializing to the case where the background spacetime is Kerr, we arrive at a canonical Hamiltonian for the test-body limit of the relativistic spinning two-body problem, valid to quadrupolar order in the test body's multipole expansion.  Our results complement those of Refs.~\cite{Steinhoff:Puetzfeld:2012,Bini:Geralico:2013, Bini:2013uwa, Bini:Faye:Geralico:2015}, which have also considered spin-squared effects for test bodies in Kerr, with one notable new feature of our results being that they allow for generic orbits and spin orientations.  Of particular interest in this respect are compact (covariant) expressions for the Riemann tensor and its couplings to the spin which are valid for generic orbits, obtained by exploiting the algebraic specialness of Riemann tensor in Kerr.  Other treatments of spinning test-particle motion in algebraically special spacetimes can be found e.g.\ in Refs.~\cite{Bini:Cherubini:Geralico:Jantzen:2006,Semerak:2015lnv}.

While much of our analysis and many of our intermediate results are fully covariant, our final result for the canonical Hamiltonian in Kerr depends on a choice of coordinates and a choice of tetrad.  
We find that a comparison with PN results can be relatively easily accomplished by using Boyer-Lindquist coordinates and the ``quasi-isotropic'' tetrad of \cite{Barausse:Racine:Buonanno:2009}, though we also trace the relationship between this tetrad and the one used by Carter \cite{Carter:1968,Znajek:1977} which diagonalizes the electric and magnetic components of the Riemann tensor.  Other choices of coordinates and tetrads are likely to yield other useful forms of the Hamiltonian, as in \cite{Kunst:2015tla}, which showed that numerical evolution of the (linear-in-spin) Hamiltonian system is improved by using Kerr-Schild coordinates and an associated tetrad.  We provide here all results, including the Ricci rotation coefficients and the Riemann tensor components, to explicitly compute the Hamiltonian (\ref{introH}) in Kerr for the two tetrads of \cite{Barausse:Racine:Buonanno:2009} and \cite{Carter:1968,Znajek:1977}.  [We should note that explicitly expressing $H$ requires algebraically solving Eqs.~(\ref{introH}) and (\ref{intromu}), since $\mu$ depends on $P_t$, but this is easily done working perturbatively in the test spin; see (\ref{quadHamil}).]

While our Hamiltonian is valid only to zeroth order in the mass ratio and to spin-squared/quadrupolar order in the test body's multipole expansion, it is valid to all PN orders and to all orders in the spin parameter of the Kerr black hole.  Expanding the Hamiltonian in powers of the Kerr spin and in PN orders allows us to make comparisons with the test-mass limits of the results of high-order PN calculations, notably, those of Refs.~\cite{Porto:Rothstein:2008:2,Steinhoff:2008ji,Hergt:2008jn,Hergt:etal:2010,Bohe:etal:2015,Levi:Steinhoff:2015:1} for next-to-leading-order spin-squared couplings, Refs.~\cite{Hartung:2011te,Hartung:Steinhoff:Schafer:2012,Marsat:2012fn,Bohe:2012mr,Levi:Steinhoff:2015:2} for next-to-next-to-leading-order spin-orbit interactions, and Refs.~\cite{Levi:Steinhoff:2014:2, Marsat:2014xea, Vaidya:2014kza, Hergt:2007ha, Hergt:2008jn} for leading-order couplings at third- and fourth-orders in the spins.  We find complete agreement with the test-mass limits of all available (complete) PN results.  We remark that the full finite-mass-ratio PN results for the leading-PN-order spin couplings for binary black holes, through fourth order in the spins, can all be ``deduced'' from the results in the test-mass limit.

While our final spinning test-body Hamiltonian is expressed in terms of canonical variables defined by the Newton-Wigner SSC, we provide the explicit translation into variables defined by other SSCs, and in particular by the more physically motivated ``covariant'' or Tulczyjew SSC \cite{Tulczyjew:1959, Fokker:1929, Dixon:1979}.  Future work in developing effective Hamiltonians for the spinning two-body problem is likely to benefit from a detailed analysis of how to expound upon this translation---relating different definitions of position, momentum, spin, and quadrupole variables---with explicit connections to the definitions used in other approaches to the spinning two-body problem, including (i) the effective action approaches to spin effects in PN theory (see e.g.\ \cite{Levi:Steinhoff:2015:1,Marsat:2014xea}), (ii) self-force calculations and their uses in determining EOB potentials (see e.g.\ \cite{Damour:SFEOB,Barausse:etal:SF,Bini:Damour:Geralico:2015}), and (iii) extracting appropriate measures of the mass, spin, and other multipole moments of black holes and fluid bodies in numerical relativity simulations (see e.g.\ \cite{OwenThesis}).

We begin in Sec.~\ref{sec:action} by discussing constrained action principles for a spinning test body, summarizing how a formulation in terms of generic-SSC variables is related to one in terms of covariant-SSC variables.  Section \ref{sec:shift} applies results from bitensor calculus to derive the transformation properties summarized in Sec.~\ref{sec:action}.  We use the Newton-Wigner SSC to achieve canonical variables in Sec.~\ref{sec:ham}.  We specialize to a spin-induced quadrupole and decompose the couplings to the Weyl/vacuum-Riemann tensor in terms of its electric and magnetic parts in Sec.~\ref{sec:quad}.  We specialize to the Kerr spacetime and its algebraically special Riemann tensor in Sec.~\ref{sec:kerr}, and we collect the necessary results and perform the PN expansion in Sec.~\ref{sec:PN}.  We conclude in Sec.~\ref{sec:discussion}.

\section{Equations of motion and action principles}\label{sec:action}

The motion of an extended test body in curved spacetime is described, in a multipolar approximation, by the Mathisson-Papapetrou-Dixon (MPD) equations \cite{Mathisson:1937, Papapetrou:1951pa, Dixon:1979}, which are given to quadrupolar order by
\begin{align}\label{eq:MPDp}
\frac{Dp^\mu}{d s}&=-\frac{1}{2}R^\mu{}_{\nu\alpha\beta}\dot z^\nu S^{\alpha\beta}-\frac{1}{6}\nabla^\mu R_{\nu\rho\alpha\beta} J^{\nu\rho\alpha\beta},\phantom{\Bigg|}
\\*\label{eq:MPDS}
\frac{DS^{\mu\nu}}{d s}&=2p^{[\mu}\dot z^{\nu]}+\frac{4}{3}R^{[\mu}{}_{\rho\alpha\beta}J^{\nu]\rho\alpha\beta},
\end{align}
where $p^\mu$ is the linear momentum vector, $S^{\mu\nu}$ is the antisymmetric angular momentum (or spin) tensor, and $J^{\mu\nu\alpha\beta}$ is the quadrupole tensor, all of which are tensors defined along a representative worldline $z^\mu( s)$ with tangent $\dot z^\mu=dz^\mu/d s$, where $ s$ is an arbitrary parameter.  
Our sign convention for the Riemann tensor is given by $2\nabla_{[a}\nabla_{b]}V_c=R_{abc}{}^dV_d$, and we use the $(-,+,+,+)$ metric signature.
The quadrupole $J^{\mu\nu\alpha\beta}$ may depend on certain internal degrees of freedom of the body, or (as in the case of a spin-induced quadrupole, or an adiabatic tidally induced quadrupole) it may be determined by only $p^\mu$, $S^{\mu\nu}$, and the local geometry along $z^\mu$.  In the latter case, Eqs.~(\ref{eq:MPDp}) and (\ref{eq:MPDS}) completely determine the evolution of $p^\mu$ and $S^{\mu\nu}$ along a given worldline $z^\mu$; however, they do not single out a choice of worldline. 

A fully determined system for evolving $p^\mu$, $S^{\mu\nu}$, and $z^\mu$ can be obtained by enforcing a spin supplementary condition (SSC), of the form $S_{\mu\nu}f^\nu = 0$.  This corresponds to demanding that the body have a vanishing mass dipole about $z^\mu$ in the local Lorentz frame defined by a timelike vector field $f^\mu$.  The worldline $z^\mu$ follows the body's center of mass as measured in the frame of $f^\mu$.  

The most common and physically sensible choice for the SSC is to use the body's rest frame, i.e.\ $f^\mu=p^\mu$, yielding the ``covariant'' (or Tulczyjew \cite{Tulczyjew:1959, Fokker:1929, Dixon:1979}) SSC: 
\begin{equation}
\tilde S_{\mu\nu}\tilde p^\nu=0,
\end{equation}
where we denote quantities defined by the covariant SSC with a tilde.  We also later insert tildes on indices for the tangent space at the point $\tilde z$, as in $\tilde S_{\tilde\mu\tilde\nu}\tilde p^{\tilde\nu}=0$, to distinguish them from unadorned indices for the tangent space at the point $z$, but we avoid the clutter of tilded indices unless it is necessary.  Another useful choice, due to its utility in achieving 3D canonical variables, is the class of Newton-Wigner SSCs \cite{Pryce:1935, Pryce:1948, Newton:Wigner:1949}, defined in terms of an arbitrary unit timelike vector field $e_0{}^\mu$ by
\begin{equation}\label{iiNW}
S_{\mu\nu}\left(\frac{p^\nu}{\sqrt{-p^2}}+e_0{}^\nu\right)=0.
\end{equation}

We discuss first in Sec.~\ref{sec:cov_action} an explicit action principle for the quadrupolar MPD equations which enforces the covariant SSC.  In Sec.~\ref{sec:gen_action}, we discuss how to generalize to an action which enforces a generic SSC, and we follow the change of variables that relates quantities defined by the covariant SSC to those defined by a generic SSC.  In Secs.~\ref{sec:action}-\ref{sec:shift}, we use unadorned symbols for quantities defined by a generic SSC, and these become those defined by a Newton-Wigner SSC in Sec.~\ref{sec:ham}, with tildes denoting covariant-SSC quantities throughout.

\subsection{Action for the covariant SSC}\label{sec:cov_action}

An explicit action functional which yields the quadrupolar MPD equations (\ref{eq:MPDp}, \ref{eq:MPDS}) while enforcing the covariant SSC is given by \cite{Steinhoff:2014, Steinhoff:Schafer:2009:2}
\begin{equation}\label{eq:Taction}
\mathcal S[\tilde p,\tilde S,\tilde z,\tilde \Lambda]=\int d s\left[\tilde p_\mu \dot {\tilde z}^\mu+\frac{1}{2}{\tilde S}_{\mu\nu}\tilde \Omega^{\mu\nu}-\tilde H_D\right].
\end{equation}
The independent degrees of freedom to be varied here are the momentum $\tilde p^\mu$, the spin $\tilde S^{\mu\nu}$, the worldline $\tilde z^\mu$, and a ``body-fixed'' orthonormal tetrad $\tilde \Lambda_A{}^\mu$ along the worldline, satisfying $\eta^{AB}\tilde \Lambda_A{}^\mu\tilde  \Lambda_B{}^\nu=g^{\mu\nu}$, from which the (antisymmetric) angular velocity tensor $\tilde \Omega^{\mu\nu}$ is defined as
\begin{equation}\label{eq:Omega}
\tilde \Omega^{\mu\nu}=\tilde \Lambda_{A}{}^{\mu}\frac{D\tilde \Lambda^{A\nu}}{d s}.
\end{equation}
The ``Dirac Hamiltonian'' $\tilde H_D$ consists only of constraints (not involving derivatives) with Lagrange multipliers, and the action (\ref{eq:Taction}) is thus reparametrization-invariant:
\begin{equation}
\tilde H_D=2\tilde \chi^\mu \tilde S_{\mu\nu} \frac{\tilde p^\nu}{\sqrt{-\tilde p^2}}+\frac{\lambda}{2}\left[\tilde p^2+\tilde {\mathcal M}^2(\tilde p,\tilde S,\tilde z)\right].
\end{equation}
The Lagrange multiplier $\tilde \chi^\mu$ enforces the covariant SSC, $\tilde S_{\mu\nu}\tilde p^\nu=0$, while the Lagrange multiplier $\lambda$ enforces the ``mass-shell constraint'', $\tilde p^2=-\tilde{\mathcal M}^2$.  The ``dynamical mass'' function $\tilde {\mathcal M}(\tilde p,\tilde S,\tilde z)$ includes, in addition to rest-mass or other internal energy contributions, couplings between the body's multipoles and the background spacetime curvature.  Taking $\tilde {\mathcal M}$ to depend on $\tilde z$ only through the metric and the Riemann tensor evaluated at $\tilde z$ leads to the quadrupolar MPD equations (\ref{eq:MPDp}, \ref{eq:MPDS}), with the quadrupole given by
\begin{equation}\label{eq:Riem_quad}
\tilde J^{\mu\nu\alpha\beta}=\frac{3\tilde p_\rho\dot{\tilde z}^{\rho}}{\tilde p^2}\frac{\partial\tilde {\mathcal M}^2}{\partial R_{\mu\nu\alpha\beta}}.
\end{equation}
We will return in Sec.~\ref{sec:quad} to discuss the specific form of $\tilde {\mathcal M}$ which corresponds to a spin-induced quadrupole, but for now we leave it as a general function of $\tilde p$, $\tilde S$, and the metric and the Riemann tensor at $\tilde z$. 

That the variation of the action (\ref{eq:Taction}) yields the MPD equations (\ref{eq:MPDp}, \ref{eq:MPDS}) with (\ref{eq:Riem_quad}) is shown in Appendix \ref{sec:vary}.  The MPD equations and the covariant SSC can then be used to solve for the relationship between the tangent $\dot {\tilde z}^\mu$ and the momentum $\tilde p^\mu$, thus yielding complete evolution equations for $\tilde p^\mu$, $\tilde S^{\mu\nu}$, and $\tilde z^\mu$ \cite{Ehlers:Rudolph:1977, Steinhoff:Puetzfeld:2012,Marsat:2014xea}.  One finds, however, that the Lagrange multiplier $\tilde \chi^\mu$ cannot be eliminated from the equation of motion for$\phantom{\Big|}\tilde \Lambda_A{}^\mu$.  This corresponds to a residual freedom to choose the timelike component $\tilde \Lambda_0{}^\mu$ of the tetrad \cite{Steinhoff:2015}.  A consistent and physically sensible choice is $\tilde \Lambda_0{}^\mu= \tilde p^\mu/\sqrt{-\tilde p^2}$; see also \cite{Hanson:Regge:1974}.

\subsection{Action for a generic SSC}\label{sec:gen_action}

Having in hand the action (\ref{eq:Taction}) which yields the MPD equations while enforcing the covariant SSC, we now turn to generalizing this action to accommodate an arbitrary SSC.  We will find, following Refs.~\cite{Levi:Steinhoff:2015:1,Steinhoff:2015}, that this can be accomplished by a judicious change of variables in the action (\ref{eq:Taction}).  We arrive at a new action which yields the same form (\ref{eq:MPDp}, \ref{eq:MPDS}) of the MPD equations for moments $ p$ and $ S$ along the worldline $ z$ defined by a generic SSC, but which entails additional curvature couplings not present/relevant for the case of the covariant SSC, which modify the relationship between the quadrupole $ J$ (or the effective dynamical mass ${\mathcal M}$) and $ p$, $S$ and $z$.

For the first step of the change of variables, following \cite{Levi:Steinhoff:2015:1}, we transform the covariant-SSC tetrad $\tilde \Lambda_A{}^\mu$ into a new (intermediate) tetrad $\bar\Lambda_A{}^\mu$ by applying a local Lorentz transformation $L^\mu{}_\nu$ which boosts the direction of the momentum $\tilde p^\mu$ into an arbitrary unit timelike vector $v^\mu$:
\begin{equation}\label{eq:Lambdabar}
\bar\Lambda_A{}^\mu=L^\mu{}_\nu\tilde \Lambda_A{}^\nu,\quad L^\mu{}_\nu=\delta^\mu_\nu- \frac{2 v^\mu \tilde p_\nu}{\sqrt{-\tilde p^2}}+\sqrt{-\tilde p^2}\frac{\omega^\mu\omega_\nu}{-\tilde p_\rho\omega^\rho},
\end{equation}
where $\omega^\mu = \tilde p^\mu/\sqrt{-\tilde p^2} + v^\mu$.  As shown in \cite{Levi:Steinhoff:2015:1}, if this is accompanied by the following transformation of the spin tensor,
\begin{equation}\label{eq:Sbar}
\bar S^{\mu\nu}=\tilde S^{\mu\nu}+2 \tilde p^{[\mu}\tilde\xi^{\nu]},\quad \tilde\xi^\mu=-\frac{\tilde S^{\mu\nu}v_\nu}{-\tilde p_\rho\omega^\rho}=\frac{\bar S^{\mu\nu}\tilde p_\nu}{-\tilde p^2}
\end{equation}
then the rotational kinematic term in the action transforms according to
\begin{equation}\label{eq:Omegabar}
\frac{1}{2}\tilde S_{\mu\nu}\tilde \Omega^{\mu\nu}=\frac{1}{2}\bar S_{\mu\nu}\bar\Omega^{\mu\nu}-\tilde\xi^\mu\frac{D\tilde p_\mu}{d s},
\end{equation}
where $\bar\Omega^{\mu\nu}=\bar\Lambda_A{}^\mu\dfrac{D\bar\Lambda^{A\nu}}{ds}$.  From its definition (\ref{eq:Sbar}), and from $\tilde S_{\mu\nu}\tilde p^\nu=0$, the new spin tensor $\bar S^{\mu\nu}$ satisfies the new SSC $\bar S_{\mu\nu} \omega^\nu=0$.  If the original tetrad satisfied $\tilde \Lambda_0{}^\mu= \tilde p^\mu / \sqrt{-\tilde p^2}$, then the new tetrad will have $\bar\Lambda_0{}^\mu= v^\mu$, and the new SSC will read
\begin{equation}\label{eq:SGC}
\bar S_{\mu\nu}\left( \frac{\tilde p^\nu}{\sqrt{-\tilde p^2}}+\bar\Lambda_0{}^\nu\right)\equiv\mathcal C_\mu= 0.
\end{equation}
This is the ``spin gauge constraint'' discussed by \cite{Steinhoff:2015}, in which the timelike component $\bar\Lambda_0{}^\mu$ of the body-fixed tetrad plays the role of a gauge field parametrizing a generic choice of SSC defined by $\mathcal C_\mu= 0$.  We obtain the ``spin gauge invariant'' action functional presented in \cite{Steinhoff:2015} by using (\ref{eq:Sbar}) and (\ref{eq:Omegabar}) in (\ref{eq:Taction}) and modifying the $\chi$ constraint to match (\ref{eq:SGC}):
\begin{align}\label{eq:SGIaction}
&\mathcal S[\tilde p,\bar S,\tilde z,\bar\Lambda]=
\nonumber\\*
&\qquad\quad\int d s\left[\tilde p_\mu \dot {\tilde z}^\mu+\frac{1}{2}\bar S_{\mu\nu}\bar\Omega^{\mu\nu}-\frac{\bar S^{\mu\nu}\tilde p_\nu}{-\tilde p^2}\frac{D\tilde p_\mu}{d s}-\bar H_D\right],
\nonumber\\*
&\quad\bar H_D=\bar\chi^\mu \mathcal C_\mu+\frac{\lambda}{2}(\tilde p^2+\tilde {\mathcal M}^2),
\end{align}
where the original covariant-SSC dynamical mass $\tilde {\mathcal M}$, given as a function of the original spin $\tilde S^{\mu\nu}$, is expressed in terms of the new spin $\bar S^{\mu\nu}$ via $\tilde S^{\mu\nu}=\tilde {\mathcal P}^\mu_\alpha \tilde {\mathcal P}^\nu_\beta\bar S^{\alpha\beta}$, which follows from (\ref{eq:Sbar}), where $\tilde {\mathcal P}^\mu_\alpha=\delta^\mu_\alpha-\tilde  p^\mu \tilde  p_\alpha / \tilde p^2$ is the projector orthogonal to $\tilde p^\mu$.  The action (\ref{eq:SGIaction}) can also be obtained by a ``minimal coupling to gravity'' of the one derived in the context of special relativity in \cite{Steinhoff:2015}.

In the case of flat spacetime, Ref.~\cite{Steinhoff:2015} demonstrated that $\mathcal C_\mu$ is a first class constraint, and thus a generator of infinitesimal gauge transformations, and that the action (\ref{eq:SGIaction}) is invariant under these ``spin gauge transformations''.  These transformations induce infinitesimal Lorentz transformations of the tetrad $\bar\Lambda_0{}^\mu$ and corresponding shifts, $\bar S^{\mu\nu}\to\bar S^{\mu\nu}+2\tilde p^{[\mu}\tilde\xi^{\nu]}$, of the spin, similar to (\ref{eq:Lambdabar}) and (\ref{eq:Sbar}), while leaving the momentum $\tilde p^\mu$ and the worldline $\tilde z^\mu$ invariant.  It is important to note that the worldline $\tilde z^\mu$ here corresponds to the worldline defined by the covariant SSC, $\tilde S_{\mu\nu}\tilde p^\nu=0$, and by the MPD equations for $\tilde p^\mu$ and $\tilde S^{\mu\nu}$.  It is not the worldline defined by the generic SSC $\mathcal C_\mu=\bar S_{\mu\nu}(\tilde p^\nu / \sqrt{-\tilde p^2}+\bar\Lambda_0{}^\nu)=0$ and the MPD equations for $\tilde p^\mu$ and $\bar S^{\mu\nu}$; the equations of motion for $\tilde p^\mu$ and $\bar S^{\mu\nu}$ resulting from the action (\ref{eq:SGIaction}) are in fact not the MPD equations. The action (\ref{eq:SGIaction}) would yield the MPD equations if the $D\tilde{p}_\mu/ds$ term were removed.

We can transform the action (\ref{eq:SGIaction}) into a form which does yield the MPD equations by making further changes of variables, including a shift of the worldline to that defined by the new generic SSC.  We will see that the necessary worldline shift, from $\tilde z( s)$ to a new worldline $z( s)$, to quadratic order in the spin, is given by moving a unit interval along the affinely parametrized geodesic whose initial tangent is the vector $\tilde\xi$ at $\tilde z$ given by (\ref{eq:Sbar}).  In other words, $z$ is the ``exponential map'' of $\tilde\xi$ at $\tilde z$,
\begin{equation}\label{eq:expdeltaz}
z=\exp_{\tilde z}\tilde\xi,\qquad\tilde\xi^{\tilde \mu}=\frac{\bar S^{\tilde \mu\tilde \nu}\tilde p_{\tilde \nu}}{-\tilde p^2},
\end{equation}
and $\tilde\xi^{\tilde \mu}$ is the ``deviation vector'' at $\tilde z$ pointing to $z$.  Here, we have inserted tildes on indices for the tangent space at $\tilde z$ to distinguish them from unadorned indices for the tangent space at $z$.  We can then define a new tetrad $\Lambda_A{}^{\mu}$ and spin $ S^{\mu\nu}$ at $ z$ by parallel transporting $\bar\Lambda_A{}^{\tilde \mu}$ and $\bar S^{\tilde \mu\tilde \nu}$ along the geodesic from $\tilde z$:
\begin{align}\label{eq:Lambdatildetilde}
\Lambda_A{}^{\mu}&=g^{\mu}{}_{\tilde \mu}\bar\Lambda_A{}^{\tilde \mu}=g^{\mu}{}_{\tilde \mu} L^{\tilde \mu}{}_{\tilde \nu}\tilde \Lambda_A{}^{\tilde \nu},
\\\label{eq:Stildetilde}
S^{\mu\nu}&=g^{\mu}{}_{\tilde \mu} g^{\nu}{}_{\tilde \nu}\bar S^{\tilde \mu\tilde \nu}
=g^{\mu}{}_{\tilde \mu} g^{\nu}{}_{\tilde \nu}\left(\tilde S^{\tilde \mu\tilde \nu}+2\tilde p^{[\tilde \mu}\tilde\xi^{\tilde \nu]}\right),
\end{align}
where $g^{\mu}{}_{\tilde \mu}(z,\tilde z)$ is the parallel propagator \cite{Synge, Poisson:Pound:Vega:2011} along the geodesic from $\tilde z$ to $ z$, and where the second equalities have used (\ref{eq:Lambdabar}) and (\ref{eq:Sbar}) to relate back to covariant-SSC quantities.  Finally, in order to obtain a canonical form for the action (and one which yields the MPD equations), we will find that we must transform to a new momentum $ p^{\mu}$ at $ z$ according to
\begin{align}\label{eq:ptildetilde}
p^{\mu}&=g^\mu{}_{\tilde \mu}\left(\tilde p^{\tilde \mu}-\frac{1}{2}R^{\tilde \mu}{}_{\tilde \nu\tilde \alpha\tilde \beta}\bar S^{\tilde \alpha\tilde \beta}\tilde\xi^{\tilde \nu}+\frac{1}{2}R^{\tilde \mu}{}_{\tilde \alpha\tilde \nu\tilde \beta}\tilde  p^{\tilde \nu}\tilde\xi^{\tilde \alpha} \tilde\xi^{\tilde \beta}\right)
\nonumber\\
&=g^{\mu}{}_{\tilde \mu}\left(\tilde p^{\tilde \mu}-\frac{1}{2}R^{\tilde \mu}{}_{\tilde \nu\tilde \alpha\tilde \beta} \tilde S^{\tilde \alpha\tilde \beta}\tilde\xi^{\tilde \nu}-\frac{1}{2}R^{\tilde \mu}{}_{\tilde \alpha\tilde \nu\tilde \beta}\tilde  p^{\tilde \nu}\tilde\xi^{\tilde \alpha} \tilde\xi^{\tilde \beta}\right).
\end{align}
With these transformations, as shown in the following section, the action (\ref{eq:SGIaction}) becomes
\begin{align}\label{eq:Gaction}
&\mathcal S[p,S,z,\Lambda]=
\\*\nonumber
&\qquad\quad\int d s\left[ p_{\mu} \dot { z}^{\mu}+\frac{1}{2} S_{\mu\nu}\Omega^{\mu\nu}-H_D+\mathcal O(S^3)\right],
\\*
&\quad H_D=\chi^{\mu} S_{\mu\nu} \left(\frac{ p^{\nu}}{\sqrt{- p^2}}+\Lambda_0{}^{\nu}\right)+\frac{\lambda}{2}( p^2+{\mathcal M}^2),
\nonumber\\*\label{mcM2}
&\quad{\mathcal M}^2=\tilde {\mathcal M}^2- R_{\mu\nu\alpha\beta}p^\mu\xi^\nu(  S^{\alpha\beta}+ p^\alpha \xi^\beta),
\end{align}
where $\Omega^{\mu\nu}=\Lambda_A{}^\mu \dfrac{D\Lambda^{A\nu}}{d s}$ and
\begin{equation}\label{eq:deltatildez}
\xi^\mu=-\frac{ S^{\mu\nu} p_\nu}{- p^2},
\end{equation}
which is the deviation vector at $ z$ pointing to $\tilde z$, at least to quadratic order in spin, as $\xi^{\mu}=-g^{\mu}{}_{\tilde \mu} \tilde\xi^{\tilde \mu}+\mathcal O(S^3)$.  The original dynamical mass $\tilde {\mathcal M}$ is expressed in terms of the new variables, to $\mathcal O(S^2)$ accuracy, by using the same functional form of $\tilde {\mathcal M}$ as for the original covariant-SSC variables but with the spin replaced by its projection ${\mathcal P}^\mu_\alpha{\mathcal P}^\nu_\beta S^{\alpha\beta}$ orthogonal to $p^\mu$,
where ${\mathcal P}^\mu_\nu=\delta^\mu_\nu-p^\mu p_\nu/p^2$.
As shown in Appendix \ref{sec:vary}, the equations of motion resulting from the action (\ref{eq:Gaction}) are the MPD equations (\ref{eq:MPDp}, \ref{eq:MPDS})  [$+\mathcal O(S^3)$] with the quadrupole given by
\begin{align}\label{JmcM}
& J^{\mu\nu\alpha\beta}=\frac{3p_\rho\dot{z}^{\rho}}{p^2}\frac{\partial {\mathcal M}^2}{\partial R_{\mu\nu\alpha\beta}}
\\\nonumber
&=g^\mu{}_{\tilde\mu}g^\nu{}_{\tilde\nu}g^\alpha{}_{\tilde\alpha}g^\beta{}_{\tilde\beta}\tilde J^{\tilde\mu\tilde\nu\tilde\alpha\tilde\beta}
\\\nonumber
&\;\;- \frac{3p_\rho\dot{z}^{\rho}}{p^2}\Big( p^{[\mu}\xi^{\nu]}p^{[\alpha}\xi^{\beta]}
+p^{[\mu}j^{\nu]\alpha\beta}+p^{[\alpha}j^{\beta]\mu\nu} \Big)
+\mathcal O(S^3) , \\\nonumber
&j^{\nu\alpha\beta} = \frac{1}{2}  \left( \xi^\nu S^{\alpha\beta} - \xi^{[\nu} S^{\alpha\beta]} \right).
\end{align}
The contributions involving $\xi^\mu$ arise from the shift (\ref{mcM2}) of the effective dynamical mass, which introduces new curvature couplings arising from the use of a generic SSC rather than the covariant SSC.  These couplings vanish (on the constraint surface) for the case of the covariant SSC, and one can see that the generic action (\ref{eq:Gaction}) reduces to the covariant-SSC action (\ref{eq:Taction}) when the gauge field $\Lambda_0{}^\mu$ is taken to be $p^\mu/\sqrt{-p^2}$. These results (unlike those in Sec.~\ref{sec:quad}) are valid for both vacuum and nonvacuum spacetimes.

The new curvature couplings in (\ref{mcM2}) arise here from the transformations (\ref{eq:Stildetilde}) and (\ref{eq:ptildetilde}) of $p^\mu$ and $S^{\mu\nu}$, and these arise, as shown in the following section, from demanding that the transformation from the covariant-SSC action (\ref{eq:Taction}) to the generic-SSC action (\ref{eq:Gaction}) preserves the canonical forms of the kinematic terms (the terms with $s$-derivatives).  This coincides with ensuring that the generic action (\ref{eq:Gaction}) also yields the MPD equations (\ref{eq:MPDp}, \ref{eq:MPDS}), as is shown in Appendix \ref{sec:vary}, which provides a physical justification of the action (\ref{eq:Gaction}) via the derivation of the MPD equations from stress-energy conservation \cite{Dixon:1979}.

For further insights into the transformation laws (\ref{eq:Stildetilde}) and (\ref{eq:ptildetilde}), we can note: the final expressions of (\ref{eq:Stildetilde}) and (\ref{eq:ptildetilde}) for $ p^\mu$ and $ S^{\mu\nu}$ are the results of solving the MPD equations along the geodesic connecting $\tilde z$ to $ z$, with $\tilde p^{\tilde\mu}$ and $\tilde S^{\tilde\mu\tilde\nu}$ as initial data at $\tilde z$, through $\mathcal O(S^2)$.  We can also note: the holonomy of the MPD equations around a loop of size $S/p$ is the identity map through $\mathcal O(S^2)$; see Eq.~(4.14) of Ref.~\cite{FNSV} with $\kappa=1/2$.  In both of these statements, the quadrupole terms in the MPD equations do not contribute at the stated orders.  It seems clear that the transformations (\ref{eq:Stildetilde}) and (\ref{eq:ptildetilde}) of $p^\mu$ and $S^{\mu\nu}$ under a shift of the worldline should follow from their definitions in terms of the body's stress-energy tensor given by Dixon \cite{Dixon:1979}, and likewise for the transformation (\ref{JmcM}) of $J^{\mu\nu\alpha\beta}$.  While making this connection explicit would require a careful analysis of the role of the surfaces of integration in Dixon's definitions, our analysis of the effective action here avoids this complication.

We will use the generic action (\ref{eq:Gaction}) as our starting point in Sec.~\ref{sec:ham}, where we specialize to the Newton-Wigner SSC.  First, in Sec.~\ref{sec:shift}, we provide a derivation of how the covariant-SSC action (\ref{eq:Taction}) is transformed into the generic-SSC action (\ref{eq:Gaction}) via the transformations (\ref{eq:ptildetilde}), (\ref{eq:Stildetilde}), (\ref{eq:expdeltaz}), and (\ref{eq:Lambdatildetilde}) of $p$, $S$, $z$, and $\Lambda$, whose inverses are (\ref{ptr}), (\ref{Str}), (\ref{exp}, \ref{xisol}), and (\ref{Lambdatr}) below.

\section{Covariant shift of the worldline}\label{sec:shift}

We now show how to consider the shift of the worldline and the transformations of quantities defined along the worldline, in a manifestly covariant manner, using the language of bitensors \cite{Synge,Dixon:1979,Bailey:Israel:1980,Poisson:Pound:Vega:2011,Vines:2014}. An alternative derivation is presented in Appendix \ref{app:shift}.

\begin{figure}
\includegraphics[scale=.6]{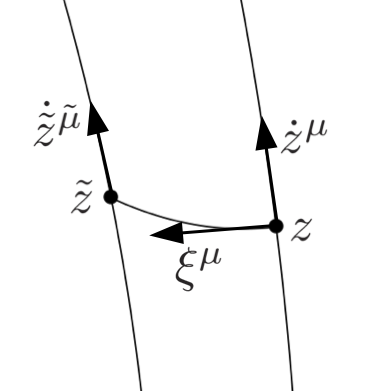}
\caption{\label{fig:xi} Along the worldline $z(s)$ defined by a generic SSC, with tangent $\dot z^\mu(s)$, we have the deviation vector field $\xi^\mu(s)$, which points (via the exponential map) to the worldline $\tilde z(s)$ defined by the covariant SSC, with tangent $\dot{\tilde z}^{\tilde\mu}(s)$.}
\end{figure}

It will be convenient to start with the worldline $z(s)$ defined by a generic SSC and shift to the worldline $\tilde z(s)$ defined by the covariant SSC.  In general, a new worldline $\tilde z(s)$ can be specified by a deviation vector field $\xi^{\mu}(s)$ along an old worldline $z(s)$, according to
\begin{equation}\label{exp}
\tilde z=\exp_z\xi\quad\Leftrightarrow\quad\xi^\mu=-\nabla^\mu\sigma(z,\tilde z),
\end{equation}
where $\sigma(z,\tilde z)$ is Synge's world function \cite{Synge,Poisson:Pound:Vega:2011}, giving half the squared proper interval along the geodesic connecting $z$ to $\tilde z$.
The point $\tilde z$ is reached by traveling a unit interval along the affinely parametrized geodesic starting at $z$ with tangent $\xi^\mu$, as in Fig.~\ref{fig:xi}.

Differentiating the second relation in (\ref{exp}),
\begin{align}
\frac{D\xi^\mu}{ds}=-\dot z^\nu\nabla_\nu\nabla^\mu\sigma-\dot{\tilde z}^{\tilde\mu}\nabla_{\tilde\mu}\nabla^\mu\sigma,
\end{align}
and solving for the tangent to $\tilde z(s)$ yields
\begin{equation}
\dot{\tilde z}^{\tilde\mu}=K^{\tilde\mu}{}_\mu\dot z^\mu+H^{\tilde\mu}{}_\mu\frac{D\xi^\mu}{ds},
\end{equation}
where
\begin{align}
H^{\tilde\mu}{}_\mu&=-\left(\nabla_{\tilde\mu}\nabla^\mu\sigma\right)^{-1}=g^{\tilde\mu}{}_\mu+\mathcal O(\xi^2),
\\*\nonumber
K^{\tilde\mu}{}_\mu&=H^{\tilde\mu}{}_\nu\nabla_\mu\nabla^\nu\sigma=g^{\tilde\mu}{}_\nu\left[\delta^\nu_\mu-\frac{1}{2}R^\nu{}_{\alpha\mu\beta}\xi^\alpha\xi^\beta+\mathcal O(\xi^3)\right]
\end{align}
are the ``Jacobi propagators'' \cite{Dixon:1979,Bailey:Israel:1980,Schattner:Trumper,Vines:2014}, with the second equalities giving their expansions in powers of the deviation vector \cite{Vines:2014}.  Thus,
\begin{equation}\label{eq:ztildedot}
\dot{\tilde z}^{\tilde\mu}=g^{\tilde\mu}{}_\mu\left(\dot z^\mu+\frac{D\xi^\mu}{ds}-\frac{1}{2}R^\mu{}_{\alpha\nu\beta}\dot z^\nu\xi^\alpha\xi^\beta+\mathcal O(\xi^3)\right),
\end{equation}
which gives the tangent to the covariant-SSC worldline $\tilde z(s)$ in terms of the generic-SSC worldline $z(s)$ and the deviation vector $\xi^\mu(s)$ along $z(s)$.

Let us take the momentum $\tilde p_{\tilde\mu}$ at $\tilde z$ to be related to the new momentum $p_\mu$ at $z$ by
\begin{equation}
\tilde p_{\tilde\mu}=g_{\tilde\mu}{}^\mu\left( p_\mu+\delta p_\mu\right),
\end{equation}
where $\delta p_\mu$ is an $\mathcal O(S^2)$ correction to be determined, anticipating that $\xi=\mathcal O(S)$.
Then,
\begin{align}
\frac{D\tilde p_{\tilde\mu}}{ds}&=g_{\tilde\mu}{}^\mu\frac{D p_{\mu}}{ds}+\left(\dot z^\nu\nabla_\nu g_{\tilde\mu}{}^\mu+\dot{\tilde z}^{\tilde\nu}\nabla_{\tilde\nu}g_{\tilde\mu}{}^\mu\right) p_\mu+\mathcal O(S^2)
\nonumber\\
&=g_{\tilde\mu}{}^\mu\left(\frac{D p_{\mu}}{ds}-R_\mu{}^\alpha{}_{\nu\beta} p_\alpha\dot z^\nu\xi^\beta+\mathcal O(S^2)\right),
\end{align}
where we have used (\ref{eq:ztildedot}) and the expansions of the derivatives of the parallel propagator \cite{Poisson:Pound:Vega:2011},
\begin{align}
\nabla_\nu g_{\tilde\mu}{}^\alpha&=-\frac{1}{2}g_{\tilde\mu}{}^\mu R_\mu{}^\alpha{}_{\nu\beta}\xi^\beta+\mathcal O(\xi^2),
\\*
\nabla_{\tilde\nu}g_{\tilde\mu}{}^\alpha&=-\frac{1}{2}g_{\tilde\mu}{}^\mu g_{\tilde\nu}{}^\nu R_\mu{}^\alpha{}_{\nu\beta}\xi^\beta+\mathcal O(\xi^2).
\end{align}
Similarly, taking the intermediate body-fixed tetrad $\bar\Lambda_A{}^{\tilde\mu}$ of (\ref{eq:Lambdabar}) at $\tilde z$ to be related to the new tetrad at $z$ by parallel transport,
$
\bar\Lambda_A{}^{\tilde\mu}=g^{\tilde\mu}{}_\mu\Lambda_A{}^\mu
$,
we have
\begin{equation}
\frac{D\bar\Lambda_A{}^{\tilde\mu}}{ds}=g^{\tilde\mu}{}_\mu\left(\frac{D\Lambda_A{}^\mu}{ds}-R^\mu{}_{\alpha\nu\beta}\Lambda_A{}^\alpha\dot z^\nu\xi^\beta+\mathcal O(S^2)\right),
\end{equation}
and thus,
\begin{align}\label{deltaOmega}
\bar\Omega^{\tilde\mu\tilde\nu}=g^{\tilde\mu}{}_\mu g^{\tilde\nu}{}_\nu\Big(\Omega^{\mu\nu}+R^{\mu\nu}{}_{\alpha\beta}\dot z^\alpha\xi^\beta+\mathcal O(S^2)\Big),
\end{align}
with $\Omega^{\mu\nu}$ and $\bar\Omega^{\tilde\mu\tilde\nu}$ as defined in (\ref{eq:Omega}) and below (\ref{eq:Omegabar}).  Finally, the intermediate spin $\bar S^{\tilde\mu\tilde\nu}$ of (\ref{eq:Sbar}) at $\tilde z$ is parallel transported into the new spin $S^{\mu\nu}$ at $z$, as in (\ref{eq:Stildetilde}).

Putting everything together, we find that the kinematic terms of the action (\ref{eq:Taction}) transform according to
\begin{widetext}
\begin{align}
\tilde p_{\tilde\mu}\dot{\tilde z}^{\tilde\mu}+\frac{1}{2}\tilde S_{\tilde\mu\tilde\nu}\tilde\Omega^{\tilde\mu\tilde\nu} 
&=\tilde p_{\tilde\mu}\dot{\tilde z}^{\tilde\mu}+\frac{1}{2}\bar S_{\tilde\mu\tilde\nu}\bar\Omega^{\tilde\mu\tilde\nu}-\frac{\bar S^{\tilde\mu\tilde\nu}\tilde p_{\tilde\nu}}{-\tilde p^2}\frac{D\tilde p_{\tilde\mu}}{ds}
\nonumber\\
&=\left( p_\mu+\delta p_\mu+\frac{1}{2}R_{\mu\nu}{}^{\alpha\beta}S_{\alpha\beta}\xi^\nu-\frac{1}{2}R_{\mu\alpha}{}^\nu{}_\beta p_\nu\xi^\alpha\xi^\beta-R_{\mu\alpha}{}^\nu{}_\beta p_\nu\xi^\alpha\frac{S^{\beta\gamma} p_\gamma}{- p^2}\right)\dot z^\mu
\nonumber\\
&\quad\;+\frac{1}{2}S_{\mu\nu}\Omega^{\mu\nu}-\left(\xi^\mu+\frac{S^{\mu\nu} p_\nu}{- p^2}\right)\frac{D p_\mu}{ds}+\frac{D}{ds}\left( p_\mu\xi^\mu\right)+\mathcal O(S^3).
\label{actiontransformation}
\end{align}
\end{widetext}
We see that we can remove the last two terms by choosing the deviation vector to be
\begin{equation}\label{xisol}
\xi^\mu=-\frac{S^{\mu\nu}p_\nu}{-p^2}.
\end{equation}
We then see that $p_\mu$ will be (covariantly) conjugate to $z^\mu$ if we choose
\begin{align}\label{ptr}
g_\mu{}^{\tilde\mu}\tilde p_{\tilde\mu}&=p_\mu+\delta p_\mu
\\*\nonumber
&=p_\mu-\frac{1}{2}R_{\mu\nu}{}^{\alpha\beta}S_{\alpha\beta}\xi^\nu-\frac{1}{2}R_{\mu\alpha}{}^\nu{}_\beta p_\nu\xi^\alpha\xi^\beta,
\end{align}
which is the inverse of (\ref{eq:ptildetilde}).  The complete transformation from generic- to covariant-SSC variables is then given by (\ref{ptr}) and
\begin{align}\label{Str}
\tilde S^{\tilde\mu\tilde\nu}&=g^{\tilde\mu}{}_\mu g^{\tilde\nu}{}_\nu(S^{\mu\nu}+2p^{[\mu}\xi^{\nu]}),
\\\label{Lambdatr}
\tilde\Lambda_A{}^{\tilde\mu}&=L_{\tilde\nu}{}^{\tilde\mu}{}g^{\tilde\nu}{}_\nu\Lambda_A{}^\nu,
\end{align}
along with the worldline shift defined by (\ref{exp}) and (\ref{xisol}).

In the end, (\ref{actiontransformation}) has become
\begin{align}
\tilde p_{\tilde\mu}\dot{\tilde z}^{\tilde\mu}+\frac{1}{2}\tilde S_{\tilde\mu\tilde\nu}\tilde\Omega^{\tilde\mu\tilde\nu} 
=p_\mu\dot z^\mu+\frac{1}{2}S_{\mu\nu}\Omega^{\mu\nu}+\mathcal O(S^3),
\end{align}
and inserting this into the covariant-SSC action (\ref{eq:Taction}) yields the generic-SSC action (\ref{eq:Gaction}), with appropriately modified Lagrange multiplier terms.  The expression (\ref{mcM2}) for the effective squared dynamical mass $\mathcal M^2=-p^2$ follows from the transformation (\ref{ptr}) of $p^\mu$ and from $\tilde{\mathcal M}^2=-\tilde p^2$.

\section{Canonical Hamiltonian}\label{sec:ham}

We now take the final form (\ref{eq:Gaction}) of the action for a generic SSC and specialize to the Newton-Wigner SSC (\ref{iiNW}), in order to obtain a Hamiltonian formulation in terms of 3D dynamical variables with canonical Poisson brackets.  

This involves a choice of an arbitrary fixed orthonormal frame or tetrad $e_a{}^\mu$ on the background spacetime, satisfying $e_a{}^\mu e_{b\mu}=\eta_{ab}$, where $\eta_{ab}$ is the Minkowski metric and is used to raise and lower the frame indices.  We write $e_a{}^\mu=(e_0{}^\mu,e_i{}^\mu)$, where the frame indices $a,b,c,\ldots$ take values $0$ for the temporal component and $i,j,k,\ldots=1,2,3$ for the spatial components.  We also use $A=(0,i)$ for the body-fixed frame indices on $\Lambda_A{}^\mu$.  We continue using Greek letters $\mu,\nu,\alpha,\beta,\ldots$ for coordinate-basis indices (though they could also have been interpreted as abstract indices up to now).  The Greek coordinate-basis indices take values $t$ for the time coordinate and $\mathrm{i},\mathrm{j},\mathrm{k},\ldots$ ($=r,\theta,\phi$, say) for the spatial coordinates, with the unitalicized font distinguishing the latter from spatial frame indices $i,j,k$.  We use frame components of tensors such as $p_a=(p_0,p_i)=e_a{}^\mu p_\mu$, to be distinguished from the coordinate-basis components $p_\mu=(p_t,p_{\mathrm{i}})$.

With this notation in order, we consider the generic-SSC action (\ref{eq:Gaction}):
\begin{align}\label{eq:noGaction}
\mathcal S&=\int ds \left[ p_{\mu} \dot {z}^{\mu}+\frac{1}{2}S_{\mu\nu}\Omega^{\mu\nu}- H_D\right],
\\
 H_D&=\chi^{a}\mathcal C_a+\frac{\lambda}{2}\Big( p^2+{{\mathcal M}}^2(p,S,z)\Big).
\nonumber
\end{align}
As discussed in \cite{Steinhoff:2015}, the spin gauge constraint,
\begin{equation}\label{EQ:SGC}
\mathcal C_a=S_{ab} \left(\frac{p^b}{\sqrt{-p^2}}+\Lambda_0{}^b\right)=0,
\end{equation}
is not itself a SSC, but it becomes a specific SSC with a specific choice of the ``gauge field'' $\Lambda_0{}^a$.  The following choices for the gauge field $\Lambda_{0}{}^{a}$
turn (\ref{EQ:SGC}) into various familiar
SSCs: 
\begin{align}
\Lambda_{0}{}^{a} &= \frac{p^a}{\sqrt{-p^2}}\quad\Rightarrow\quad S_{ab}p^{b}=0,
\\
\Lambda_{0}{}^{a} &= \frac{2p^{0}\delta_{0}^{a}-p^{a}}{\sqrt{-p^2}}\quad\Rightarrow\quad S_{a0}=0,
\\
\Lambda_{0}{}^{a} &= \delta_{0}^{a}\quad\Rightarrow\quad S_{ab}(p^{b}+\sqrt{-p^2}\delta_{0}^{b})=0. \label{eq:NWSSC}
\end{align}
The first choice represents the covariant Tulczyjew SSC \cite{Tulczyjew:1959, Fokker:1929} and the second yields the Corinaldesi-Papapetrou SSC \cite{Corinaldesi:Papapetrou:1951, Pryce:1948, Moller:1949}. The third condition (\ref{eq:NWSSC}), leading to the Newton-Wigner (NW) SSC \cite{Pryce:1935, Pryce:1948, Newton:Wigner:1949}, will be the one used here. The NW SSC allows one to formulate a canonical phase space algebra for the reduced degrees of freedom on the constraint surface, as we shall see below. In general relativity, this SSC saw a widespread use only more recently. It was employed for post-Newtonian calculations in \cite{Porto:2006, Levi:2008, Levi:Steinhoff:2015:1, Steinhoff:2008}, where \cite{Levi:2008, Levi:Steinhoff:2015:1} apply it in the Feynman rules, in the ADM canonical formulation of spin \cite{Steinhoff:2008, Steinhoff:2011}, for the test-spin Hamiltonian in \cite{Barausse:Racine:Buonanno:2009}, and at the level of the MPD equations in \cite{Lukes-Gerakopoulos:2014}. However, while Refs.~\cite{Porto:2006, Levi:2008, Barausse:Racine:Buonanno:2009} use the condition on the spin in (\ref{eq:NWSSC}), their condition on $\Lambda_{0}{}^{a}$ differs from (\ref{eq:NWSSC}).

It is useful to write the rotational kinematic term in the action in the
local frame, 
\begin{align}
S_{\mu\nu}\Omega^{\mu\nu} & =S_{\mu\nu}\Lambda_{A}{}^{a}e_{a}{}^{\mu}\frac{D(\Lambda^{Ab}e_{b}{}^{\nu})}{ds}
\nonumber\\*\label{eq:firstomega}
 & =S_{ab}\left(\Lambda_{A}{}^{a}\dot{\Lambda}^{Ab}+\omega_{\mu}{}^{ab}\dot z^{\mu}\right),
\end{align}
where dots denote the ordinary derivative $d/ds$, and 
\begin{equation}\label{rotcos}
\omega_{\mu}{}^{ab}=e^{b}{}_{\nu}\nabla_\mu e^{a\nu}
\end{equation}
are the Ricci rotation coefficients.  Choosing the NW SSC (\ref{eq:NWSSC}) removes all temporal components from
the $S\Lambda\dot\Lambda$ term (notice that also $\Lambda_{A}{}^{0}=\delta_{A}^{0}$), leaving only spatial components:
\begin{align}\label{eq:nozeros}
S_{ab}\Lambda_{A}{}^{a}\dot{\Lambda}^{Ab}=S_{ij}\Lambda^{ki}\dot{\Lambda}^{kj},
\end{align}
where we understand that the first index of $\Lambda^{ki}$ refers
to the body-fixed frame and the second one to the local frame.  Thus, the RHS of (\ref{eq:nozeros}) provides a canonical kinematic term for the physical degrees of freedom $\Lambda^{ij}$ and $S_{ij}$, and the dependent degrees of freedom $\Lambda_0{}^\mu$ and $S_{0i}$ have no kinematic terms.  The latter are fixed by the gauge choice $\Lambda_0{}^a=\delta_0^a$ and the resultant NW SSC (\ref{eq:NWSSC}), which can be solved to yield
\begin{equation}\label{S0s}
S_{0i}=\frac{S_{ij}p^j}{p^{0}+{{\mathcal M}}},
\end{equation}
having used $p^2=-{\mathcal M}^2$.  These arguments allow us to avoid the Dirac brackets for handling the constraints, which would be considerably more complicated \cite{Barausse:Racine:Buonanno:2009}.

Using (\ref{eq:firstomega}) and (\ref{eq:nozeros}) in (\ref{eq:noGaction}), we see that the action in the NW SSC has the form
\begin{equation}\label{almostthere}
\mathcal S=\int ds\left[ P_{\mu} \dot {z}^{\mu}+\frac{1}{2}S_{ij}\Lambda^{ki}\dot{\Lambda}^{kj}- H_D\right],
\end{equation}
where we have defined a new momentum,
\begin{equation}\label{capitalP}
P_{\mu}=p_\mu+\frac{1}{2}\omega_\mu{}^{ab}S_{ab},
\end{equation}
whose coordinate-basis components $P_\mu=(P_t,P_{\,\mathrm{i}})$ are canonically conjugate to the worldline coordinates $z^\mu=(t,z^{\,\mathrm{i}})$.  We refer to $P_\mu$ as the canonical momentum and to $p_\mu$ as the covariant momentum, and we work with both below.

The form (\ref{almostthere}) of the action still has unphysical degrees of freedom associated with reparametrization invariance.  We can fix these by choosing the worldline parameter to be the time coordinate, $s=t$, so that $\dot t=1$, and thus,
\begin{equation}\label{TIMEGAUGE}
P_\mu\dot z^\mu=P_t+P_{\,\mathrm{i}\,}\dot z^{\mathrm{i}}.
\end{equation}
We can then solve the mass-shell constraint $p^2=-\mathcal M^2$ for $P_t$, using (\ref{capitalP}).  This is most easily accomplished order by order in the spin, and we will discuss the solution to linear order in the following subsection and to quadratic order in Sec.~\ref{sec:PN}.

Having solved both constraints, $H_D$ vanishes, and we obtain from (\ref{almostthere}) and (\ref{TIMEGAUGE}) the final canonical form of the action,
\begin{equation}\label{finalgeneralaction}
\mathcal S=\int dt\left[P_{\,\mathrm{i}\,}\dot z^{\mathrm{i}}+\frac{1}{2}S_{ij}\Lambda^{ki}\dot{\Lambda}^{kj}- H\right],
\end{equation}
where
\begin{equation}
H(z^{\mathrm{i}},P_{\,\mathrm{i}\,},S_{ij})=-P_t.
\end{equation}
A variation of the action with respect to the dynamical variables $z^{\mathrm{i}}$, $P_{\,\mathrm{i}}$, $\Lambda^{ij}$, and $S_{ij}$ leads to the equations of motion
\begin{equation}
\dot z^{\mathrm{i}}=\frac{\partial H}{\partial P_{\,\mathrm{i}}},\quad \dot P_{\,\mathrm{i}}=-\frac{\partial H}{\partial z^{\mathrm{i}}},\quad\dot S_i=\epsilon_{ijk}\frac{\partial H}{\partial S_j}S_k,
\end{equation}
where
\begin{equation}
S_i=\frac{1}{2}\epsilon_{ijk}S_{jk}.
\end{equation}
These have the form of Hamilton's canonical equations with $H$ being the Hamiltonian. The canonical Poisson brackets for the dynamical variables $z^{\mathrm{i}}$, $P_{\,\mathrm{i}}$, and $S_{ij}$ can be ``read off'' from these equations of motion as
\begin{equation}\label{PBs}
\{z^{\mathrm{i}},P_{\,\mathrm{j}}\}=\delta^{\mathrm{i}}_{\mathrm{j}},\qquad \{S_i,S_j\}=\epsilon_{ijk}S_k,
\end{equation}
with all others vanishing.

\subsection{Hamiltonian to linear order in spin}

We can find the explicit Hamiltonian $H=-P_t$ to linear order in the spin by solving the mass shell constraint $p^2=-{\mathcal M}^2$ for $P_t$ in terms of $z^{\mathrm{i}}$, $P_{\,\mathrm{i}}$, and $S_{ij}$, using $P_\mu=p_\mu+\frac{1}{2}\omega_\mu{}^{ab}S_{ab}$ as in (\ref{capitalP}), and using the solution for $S_{0i}$ given by (\ref{S0s}).  Defining the lapse $N$, shift $N^{\mathrm{i}}$, and inverse spatial metric $\gamma^\mathrm{ij}$ of the background spacetime,
\begin{align}
N & =\frac{1}{\sqrt{-g^{tt}}},
\nonumber\\\label{lapseshift}
N^{\mathrm{i}} & =N^{2}g^{t{\mathrm{i}}}=-\frac{g^{t{\mathrm{i}}}}{g^{tt}},
\\\nonumber
\gamma^{\mathrm{ij}} & =g^\mathrm{ij}+\frac{N^\mathrm{i}N^\mathrm{j}}{N^{2}}=g^\mathrm{ij}-\frac{g^{t\mathrm{i}}g^{t\mathrm{j}}}{g^{tt}}\,,
\end{align}
we find
\begin{align}\label{linSHam}
H&=-P_t(z^{\mathrm{i}},P_{\,\mathrm{i}\,},S_{ij})
\\\nonumber
&=H_\mathrm{NS}-\frac{N}{Q}\hat P_\mu\left(\frac{\omega^{\mu ij}}{2}+\frac{\omega^{\mu 0i}\hat P^{j}}{\hat P^{0}+{m}}\right)S_{ij}+\mathcal O(S^2),
\end{align}
where
\begin{align}\label{HNS}
H_\mathrm{NS}&=NQ-N^{\mathrm{i}}P_{\,\mathrm{i}\,},
\\\nonumber
Q&=\sqrt{{m}^2+\gamma^\mathrm{ij}P_{\,\mathrm{i}}P_{\,\mathrm{j}\,}},
\\\nonumber
\hat P_\mu&=(-H_\mathrm{NS},P_{\,\mathrm{i}}),
\\\nonumber
\hat P^a&=e^{a\mu}\hat P_\mu=(\hat P^{0},\hat P^{i})=(e^{0\mu}\hat P_\mu, e^{i\mu}\hat P_\mu),
\end{align}
and where we have taken ${\mathcal M}^2=m^2+\mathcal O(S^2)$ with $m$ being a constant.

The Hamiltonian becomes somewhat simpler if we adopt the ``time gauge'' \cite{Schwinger:1963:1}, i.e.\ if we specialize the local Lorentz frame $e_a{}^\mu$ so that its timelike vector points along the direction of the time coordinate, so that $e^0{}_\mu=N\delta^t_\mu$ and also $e_a{}^t=\delta_a^0/N$. We will refer to this choice as a time-aligned tetrad from now on. This choice also implies that $P^{0}=N P^t$ and thus, from (\ref{lapseshift}) and (\ref{HNS}), that $\hat P^{0}=Q$.  It also implies that $\hat P^{i}=e^{i\mathrm{j}}P_{\,\mathrm{j}}=P^i=P_i$, which is then independent of $P_t$.  We can then write the Hamiltonian (\ref{linSHam}) as
\begin{equation}\label{linSHamTG}
H=H_\mathrm{NS}-\frac{N}{Q}\hat P^a\left(\frac{\omega_{aij}}{2}-\frac{\omega_{a0i}P_{j}}{Q+{m}}\right)S^{ij}+\mathcal O(S^2),
\end{equation}
where $\hat P^a=(Q,P^i)$, with $H_\mathrm{NS}$ and $Q$ still given by (\ref{HNS}).  This agrees with Eqs.~(4.41-45) of \cite{Barausse:Racine:Buonanno:2009} if we note $\omega_{\mu ab}=2 E_{\mu ab}$ and mind some raised and lowered indices and changes of bases.

\section{Curvature couplings at quadratic order in spin}\label{sec:quad}

At quadratic order in the spin, the action is still given by (\ref{finalgeneralaction}), with the Hamiltonian $H=-P_t$ determined by solving the mass-shall constraint $p^2=-{\mathcal M}^2$, where $P_\mu=p_\mu+\frac{1}{2}\omega_\mu{}^{ab}S_{ab}$ as in (\ref{capitalP}).  But we must now also take into account the spin-squared contributions to the effective dynamical mass $\mathcal M$, which arise both from intrinsic couplings in the covariant-SSC dynamical mass $\tilde{\mathcal M}$ and from what one might call the kinematic couplings of (\ref{mcM2}),
\begin{equation}\label{kincoup}
\mathcal M^2=\tilde{\mathcal M}^2-R_{abcd}p^a\xi^b(S^{cd}+p^c\xi^d)+\mathcal O(S^3),
\end{equation}
where 
\begin{equation}\label{mydeltaz}
\xi^a=-\dfrac{S^{ab}p_b}{-p^2}.
\end{equation}
The form of the covariant-SSC dynamical mass $\tilde{\mathcal M}$ which encodes a spin-induced quadrupole moment is given by \cite{Porto:Rothstein:2008:2, Steinhoff:2011, Steinhoff:2014}
\begin{align}\label{siq}
\tilde{\mathcal M}^2&= m^2+CR_{\tilde a \tilde b \tilde c \tilde d}\frac{\tilde{p}^{\tilde a} \tilde{p}^{\tilde c}}{-\tilde{p}^2}\tilde S^{\tilde b \tilde e}\tilde S^{\tilde d}{}_{\tilde e}+\mathcal O(S^3)
\\*
&=m^2+CR_{abcd}\frac{p^ap^c}{-p^2}\tilde S^{be}\tilde S^d{}_e+\mathcal O(S^3),
\end{align}
where $m$ and $C$ are constants, and
\begin{equation}\label{tildS}
\tilde S^{ab}=\mathcal P^a_c\mathcal P^b_dS^{cd}=S^{ab}+2p^{[a}\xi^{b]}
\end{equation}
is the projection of the spin tensor orthogonal to the momentum (which coincides with the covariant-SSC spin tensor $\tilde S^{\tilde a\tilde b}$, up to parallel transport, at the considered order).  The constant $C$ measures the body's spin-induced quadrupolar deformation response.  It is equal to 1 when the body is a black hole \cite{Porto:Rothstein:2008:2, Poisson:1997ha}, and we will see that special simplifications occur in this case. For material bodies such as neutron stars, $C$ depends on the equation of state \cite{Laarakkers:1997hb, Poisson:1997ha}.
The constant mass $m(S)$ is a function of the likewise constant spin length $S = \frac{1}{2} \sqrt{\tilde{S}^{ab} \tilde{S}_{ab}}$ and encodes the moment of inertia \cite{Steinhoff:2014}. Notice that the spin length is defined with the projected spin tensor (or with the covariant-SSC spin tensor).

The couplings to the Riemann tensor-----the kinematic couplings of (\ref{kincoup}) and the intrinsic spin-induced quadrupole coupling of (\ref{siq})-----can be better understood by using the electric/magnetic decomposition of the Weyl tensor.  This goes hand-in-hand with the decomposition of the spin tensor $S^{ab}$ in terms of a Pauli-Lubanski spin vector $s^a$ (\ref{PLSV}) and the vector $\sqrt{-p^2}\xi^a$ which encodes the mass dipole.

We restrict attention to vacuum spacetimes in four dimensions.  Then the Riemann tensor equals the Weyl tensor, and it can be decomposed into contributions from an electric part
$E^{(p)}_{ab}$ and a magnetic part $B^{(p)}_{\mu\nu}$ with respect
to a timelike vector $p^{\mu}$ \cite{Novello:1980ay,McIntosh:1994,Bonnor:1995zf}.
In a compact complex notation this reads
\begin{align}
E^{(p)}_{ab} + i B^{(p)}_{ab} &= \frac{1}{2} G_{ac}{}^{ef} R_{efbd} \frac{p^cp^d}{-p^{2}} \label{EBsym} \\*
&= (R_{acbd}+i\,^*\!R_{acbd}) \frac{p^cp^d}{-p^{2}} , \label{EB}
\end{align}
where
\begin{equation}\label{bigG}
G_{abcd} = g_{ac} g_{bd} - g_{ad} g_{bc} + i \eta_{abcd} 
\end{equation}
is the tensor which projects a 2-form onto (four times) its anti-self-dual part, where the volume form is $\eta_{\mu\nu\alpha\beta}=\sqrt{-g}\epsilon_{\mu\nu\alpha\beta}$ or $\eta_{abcd}=\epsilon_{abcd}$ with $\epsilon_{0123} = 1$, and where
${}^*\!R_{acbd}=\tfrac{1}{2}\eta_{ac}{}^{ef}R_{efbd}$ is the dual of the Riemann tensor.
The tensors $E^{(p)}_{ab}$ and $B^{(p)}_{ab}$ are orthogonal to $p^a$, and thus effectively three-dimensional, and are symmetric and trace-free, making them easier to handle than $R_{abcd}$.
The following useful relations hold,
\begin{align}
G_{ab}{}^{ef} G_{efcd} &= 4 G_{abcd} , \\*
G_{abg}{}^{e} G_{cd}{}^{gf} \frac{p_ep_f}{-p^2} &= - G_{abcd} , \label{Gpinvert} \\*
R_{abcd}+i\,^*\!R_{abcd} &= \frac{1}{2} G_{ab}{}^{ef}R_{efcd} \nonumber \\*
&= \frac{1}{2} R_{abgh} G_{cd}{}^{gh} \nonumber \\*
&= \frac{1}{8} G_{ab}{}^{ef} R_{efgh} G_{cd}{}^{gh} . \label{RGcommute}
\end{align}
Note that a proof of (\ref{Gpinvert}) can require using $\eta_{[abcd}p_{e]}=0$.  In (\ref{RGcommute}), the equality of the left and right duals of the Riemann tensor was used.
From these relations together with (\ref{EBsym}), the Riemann tensor can be recovered as the real part of
\begin{equation}
R_{abcd}+i\,^*\!R_{abcd}
= G_{ab}{}^{ef} G_{cd}{}^{gh} \frac{p_ep_g}{-p^2} \left( E^{(p)}_{fh} + i B^{(p)}_{fh} \right) .\label{recoverR}
\end{equation}

Using (\ref{EB}) along with (\ref{tildS}) allows us to express the curvature couplings in (\ref{kincoup}) and (\ref{siq}) as
\begin{align}\label{cucos}
R_{abcd}p^ap^c\tilde S^b{}_e\tilde S^{de}&=p^2E^{(p)}_{ab}s^as^b,
\\*\nonumber
R_{abcd}p^{a}\xi^{b}p^c\xi^{d} & = -p^{2}E^{(p)}_{ab}\xi^{a}\xi^{b},
\\*\nonumber
R_{abcd}p^a\xi^bS^{cd} & = 2\sqrt{-p^2}B^{(p)}_{ab}s^{a}\xi^{b}  +2 p^2E^{(p)}_{ab}\xi^{a}\xi^{b},
\end{align}
where the Pauli-Lubanski spin vector $s^{a}$ is defined as
\begin{align}
\label{PLSV}
s^{a}&=-\frac{1}{2}\eta^{abcd}\frac{p_{b}}{\sqrt{-p^2}}\tilde{S}_{cd}
\nonumber\\*
&=-\frac{1}{2}\eta^{abcd}\frac{p_{b}}{\sqrt{-p^2}}{S}_{cd}.
\end{align}
One further useful identity, which follows from (\ref{recoverR}), is
\begin{align}\label{spinorresult}
&\frac{1}{4}R_{abcd}S^{ab}S^{cd}
\\\nonumber
&=-E^{(p)}_{ab}s^as^b-2\sqrt{-p^2}B^{(p)}_{ab}s^a\xi^b-p^2E^{(p)}_{ab}\xi^a\xi^b.
\end{align}
By combining (\ref{kincoup}), (\ref{siq}), (\ref{cucos}), and (\ref{spinorresult}), we can express the total effective dynamical mass as
\begin{align}\label{masterM}
\mathcal M^2&=m^2-CE^{(p)}_{ab}s^as^b-2\sqrt{-p^2}B^{(p)}_{ab}s^a\xi^b-p^2E^{(p)}_{ab}\xi^a\xi^b
\nonumber\\
&=m^2+\frac{1}{4}R_{abcd}S^{ab}S^{cd}-(C-1)E^{(p)}_{ab}s^as^b.\;\;\;
\end{align}

The $R_{abcd}S^{ab}S^{cd}$ coupling was also considered e.g.\ in \cite{Yee:Bander:1993,d'Ambrosi:2015gsa}, but therein the prefactor is an arbitrary constant, analogous to $C$ here. However, the present derivation shows that this prefactor is actually fixed (by kinematics). As was argued in \cite{Levi:Steinhoff:2015:1}, the only nonminimal couplings which carry arbitrary coefficients should be constructed from the projected spin $\tilde{S}^{ab}$ (or the vector $s^a$). The coupling terms agree with \cite{Porto:Rothstein:2008:2} in the case of the covariant SSC.

Using (\ref{masterM}), and using $P_a=p_a+\frac{1}{2}\omega_a{}^{bc}S_{bc}$ as in (\ref{capitalP}), we can rewrite the mass-shell constraint $p^2=-{\mathcal M}^2$ as
\begin{align}\label{eq:msq}
\mu^{2}&\equiv -P^2 
\\
&= m^{2}-P^a\omega_{a}{}^{bc}S_{bc}+\frac{1}{4}\omega_{a}{}^{bc}\omega^{ade}S_{bc}S_{de}
\nonumber\\\nonumber
&\quad+\frac{1}{4}R_{abcd}S^{ab}S^{cd}-(C-1)E^{(p)}_{ab}s^as^b+\mathcal{O}(S^{3}).
\end{align}
We can then give a formal solution for the Hamiltonian as
\begin{equation}\label{muHam}
H=-P_t=N\sqrt{\mu^2+\gamma^\mathrm{ij}P_{\,\mathrm{i}}P_{\,\mathrm{j}}}-N^{\mathrm{i}}P_{\,\mathrm{i}}.
\end{equation}
This is only a formal solution because $\mu^2$ depends on $P_t$.  But because this dependence starts only at $\mathcal O(S)$, this equation can be relatively easily solved for $P_t$ order by order in the spin.  We saw the fully expanded solution to linear order in spin, for the general case in (\ref{linSHam}), and with a time-aligned tetrad, $e^0{}_\mu=N\delta^t_\mu$, in (\ref{linSHamTG}). We give the solution to quadratic order, in the case of a time-aligned tetrad, in (\ref{quadHamil}).

We conclude this section with a remark on the conserved spin length.
The action (\ref{eq:noGaction}) has a symmetry under spatial rotations of the
body-fixed frame. The corresponding Noether conserved quantity is the
spatial spin in the body-fixed frame $\Lambda_i{}^a \Lambda_j{}^b S_{ab}$. Contracting this tensor with itself,
we obtain the conserved scalar
\begin{equation}
\tilde{S}_{\mu\nu} \tilde{S}^{\mu\nu} = 2 s^a s_a \equiv 2 S^2 ,
\end{equation}
where $S$ is the conserved spin length.
It should be noted that the scalar $S_{ab} S^{ab}$ is in general not conserved.
The constant mass $m(S)$ is actually a function of $S$ and encodes the moment of inertia \cite{Steinhoff:2014}.

\section{The Kerr spacetime and its Riemann tensor}\label{sec:kerr}

We now specialize to the case where the background is the Kerr geometry, giving the vacuum spacetime around a spinning black hole with mass $M$ and angular momentum $Ma$.
The metric in Boyer-Lindquist coordinates $(t,r,\theta,\phi)$ reads
\begin{equation}
\begin{split}
ds^2&=-\left(1-\frac{2Mr}{\Sigma}\right)dt^2+\frac{\Sigma}{\Delta}dr^2+\Sigma\, d\theta^2 \\
&\quad+\frac{\Lambda}{\Sigma}\sin^2\theta\,d\phi^2-\frac{4Mar}{\Sigma}\sin^2\theta\,dt\,d\phi,
\end{split}
\end{equation}
where
\begin{equation}\label{Sigma_etc}
\begin{split}
\Sigma &= r^{2}+a^{2}\cos^{2}{\theta},\\
\Delta &= \varpi^{2}-2Mr,\\
\Lambda &= \varpi^4-\Delta \,a^2\sin^2\theta,\\
\varpi^{2} &= r^{2}+a^{2}.
\end{split}
\end{equation}

The Riemann tensor of the Kerr spacetime is algebraically special, of Petrov type D, meaning that it has two repeated principal null directions (PNDs) \cite{Penrose:Rindler:1987,Stewart:1993}.  
As follows from the decomposition of the Weyl tensor in terms of the Weyl spinor, the Riemann/Weyl tensor of any vacuum type D spacetime can be written, along with its dual, in the compact complex form
\begin{align}\label{RiR}
R_{acbd}+i\,^*\!R_{acbd}&=
-\psi\left[G_{acbd}+\frac{3}{4} (G\tau)_{ac} (G\tau)_{bd}\right].
\end{align}
Here, 
$\psi$ is a coordinate-invariant complex scalar amplitude, $G_{acbd}$ is as in (\ref{bigG}), and
\begin{equation}
(G\tau)_{ab} = G_{ab}{}^{cd} \tau_{cd} = 2 (\tau_{ab}+i\chi_{ab}),
\end{equation}
where $\tau_{ab}$ is the real simple bivector/2-form spanned by the two PNDs (with $\tau_{ab}\tau^{ab}=2$, with the sign of $\tau_{ab}$ being inconsequential), and $\chi_{ab}={}^*\!\tau_{ab}=\tfrac{1}{2}\eta_{ab}{}^{cd}\tau_{cd}$ is its dual.

It is convenient to use the orthonormal tetrad $e_a{}^\mu$ on Kerr introduced by Carter \cite{Carter:1968,Znajek:1977}, for which the two PNDs are the directions of $e_0{}^\mu\pm e_1{}^\mu$, given by
\begin{equation}\label{CarterTetrad}
\left(e_{a}{}^{\mu}\right)=\left(\begin{tabular}{cccc}
 \ensuremath{\dfrac{\varpi^2}{\sqrt{\Delta\Sigma}}} &  0  &  0  &  \ensuremath{\dfrac{a}{\sqrt{\Delta\Sigma}}}\\
0  &  \ensuremath{\sqrt{\dfrac{\Delta}{\Sigma}}} &  0  &  0 \\
0  &  0  &  \ensuremath{\dfrac{1}{\sqrt{\Sigma}}} &  0 \\
\ensuremath{\dfrac{a\sin\theta}{\sqrt{\Sigma}}} &  0  &  0  &  \ensuremath{\dfrac{1}{\sqrt{\Sigma}\sin\theta}}
\end{tabular}\right),
\end{equation}
with $a=(0,i)=(0,1,2,3)$ running down and $\mu=(t,\mathrm{i})=(t,r,\theta,\phi)$ running across.  We will refer to the tetrad $e_a{}^\mu$ as the curvature-aligned frame. The components of the 2-forms $\tau_{ab}$ and $\chi_{ab}$ are given in this frame by
\begin{equation}\label{tauchi}
\tau_{ab}=2\delta_{[a}^0\delta^1_{b]},\qquad \chi_{ab}=-\epsilon_{01ab}.
\end{equation}

The complex amplitude $\psi$ for Kerr is given by
\begin{equation}\label{eib}
\psi\equiv E-iB=\frac{M}{(r+ia\cos\theta)^3},
\end{equation}
where we have defined the real scalars $E$ and $B$, with signs chosen to make them both positive in the region of interest.

Using (\ref{RiR}), the electric and magnetic parts of the Riemann tensor (\ref{EB}) with respect to $p^a$ are compactly and covariantly expressed as
\begin{align}\label{EiB}
E^{(p)}_{ab}+iB^{(p)}_{ab}&=\left(R_{acbd}+i\,^*\!R_{acbd}\right)\frac{p^cp^d}{-p^2}
\\*\nonumber
&=(E-iB)\Big[-g_{ab}g_{cd}+g_{ad}g_{bc}
\\*\nonumber
&\qquad\quad\;\;-3(\tau_{ac}+i\chi_{ac})(\tau_{bd}+i\chi_{bd})\Big]\frac{p^cp^d}{-p^2}.
\end{align}
If we take the electric/magnetic decomposition with respect to the timelike component of the curvature-aligned frame $e_0{}^a$, instead of $p^a$, then the components of $E_{ab}$ and $B_{ab}$ in the curvature-aligned frame are purely spatial, symmetric, trace-free, diagonal tensors given by
\begin{align}\label{EBcaf}
E^{(e_0)}_{ab}&=-E\delta^i_a\delta^j_b(3n_in_j-\delta_{ij}),
\\*
B^{(e_0)}_{ab}&=B\delta^i_a\delta^j_b(3n_in_j-\delta_{ij}),
\end{align}
where $n_i=\delta_i^1$.

These results allow us to easily generate explicit expressions for the curvature couplings (\ref{masterM}) in Kerr, using either the fully covariant expressions (\ref{RiR}) and (\ref{EiB}) for the Riemann tensor and its electric and magnetic parts, or the particularly simple components (\ref{EBcaf}) in the curvature-aligned frame.  We will carry this to fruition for two cases of interest.  First, in Sec.~\ref{sec:covEB}, we calculate the curvature couplings in the curvature-aligned frame $e_a{}^\mu$, restricting attention to the case $C=1$.  Then, in Sec.~\ref{sec:NWEB}, we consider general values of $C$ and general tetrads, and we introduce a second tetrad $f_a{}^\mu$, given by a boost of $e_a{}^\mu$, which satisfies the time-aligning condition discussed above (\ref{linSHamTG}).

\subsection{Curvature couplings for a test black hole in the curvature-aligned frame}\label{sec:covEB}

Here we write out the $C=1$ curvature couplings in the dynamical mass (\ref{masterM}) in terms of the curvature-aligned-frame components of the spin tensor $S_{ab}$.  We can exploit an identity analogous to (\ref{spinorresult}), using the decomposition (\ref{recoverR}) but with $p^a\to e_0{}^a$, to express the dynamical mass (\ref{masterM}) with $C=1$ as
\begin{align}\label{eq:Ce1}
\mathcal M^2&=m^2+\frac{1}{4}R_{abcd} S^{ab} S^{cd}
\\\nonumber
&=m^2-E^{(e_0)}_{ab}\hat s^a\hat s^b-2B^{(e_0)}_{ab}\hat s^a\hat \xi^b+E^{(e_0)}_{ab}\hat \xi^a\hat\xi^b,
\end{align}
where we have defined vectors $\hat s^a$ and $\hat \xi^a$ analogous to (\ref{mydeltaz}) and (\ref{PLSV}) but with $p^a\to e_0{}^a$,
\begin{align}
\hat s^a&=-\frac{1}{2}\eta^{abcd}e_{0b} S_{cd}=\frac{1}{2}\delta^a_i\epsilon_{ijk}S_{jk}=\delta^a_i S_i,
\\*
\hat\xi^a&=-S^{ab}e_{0b}=\delta^a_i S_{0i},
\end{align}
whose frame components are purely spatial and are given directly by the frame components of the spin tensor $S_{ab}$.  Using (\ref{eq:Ce1}) with (\ref{EBcaf}) then gives the remarkably simple result
\begin{equation}
\mathcal M^2=m^2+(3n_in_j-\delta_{ij})\Big[E( S_i S_j- S_{0i} S_{0j})-2BS_iS_{0j}\Big] ,
\end{equation}
in the curvature-aligned frame with $C=1$.

Recall that the temporal components $S_{0i}$ of the spin tensor are determined by solving the SSC, which gives them in terms of the spatial components $S_i=\tfrac{1}{2}\epsilon_{ijk}S_{jk}$ and the momentum $p^a=(p^0,p^i)$ with $p^2=-\mathcal M^2$.  For the NW SSC, as in (\ref{S0s}), we have
\begin{equation}\label{NWSSCSi0}
S_{0i}=\frac{\epsilon_{ijk}p_jS_k}{p^0+\mathcal M}.
\end{equation}
For the covariant SSC, $\tilde S_{ab}\tilde p^b=0$, we would have
\begin{equation}\label{covSSCSi0}
\tilde S_{0i}=\frac{\epsilon_{ijk}\tilde p_j\tilde S_k}{\tilde p^0}.
\end{equation}

\subsection{Curvature couplings for general test bodies}\label{sec:NWEB}

\subsubsection{In a general frame}

Using (\ref{EiB}), and noting that $p_as^a=p_a\xi^a=s_a\xi^a=0$, the curvature couplings in the dynamical mass (\ref{masterM}) can be expressed as
\begin{align}\label{gfccs}
-CE^{(p)}_{ab}s^as^b&=-C\big[E(s_as^a-3\tau_s^2+3\chi_s^2)
-6B\tau_s\chi_s\big],
\nonumber\\
-2\mathcal MB^{(p)}_{ab}s^a\xi^b&=6E(\tau_s\chi_\xi+\tau_\xi\chi_s)-6B(\tau_s\tau_\xi-\chi_s\chi_\xi),\phantom{\bigg|}
\nonumber\\
\mathcal M^2E^{(p)}_{ab}\xi^a\xi^b&=E(\mathcal M^2 \xi_a\xi^a-3\tau_\xi^2+3\chi_\xi^2)
-6B\tau_\xi\chi_\xi,
\end{align}
where
\begin{align}\label{sxicontrs}
\tau_s&=\tau_{ab}\frac{p^a}{\mathcal M} s^b,\qquad\qquad \chi_s=\chi_{ab}\frac{p^a}{\mathcal M}s^b,
\\\nonumber
\tau_\xi&=\tau_{ab}\frac{p^a}{\mathcal M}\mathcal M \xi^b,\qquad \quad\chi_\xi=\chi_{ab}\frac{p^a}{\mathcal M}\mathcal M \xi^b.
\end{align}
The components of the Pauli-Lubanski spin vector and the mass dipole vector,
\begin{equation}
s^a=-\frac{1}{2}\eta^{abcd}\frac{p_b}{\mathcal M}S_{cd},\qquad \mathcal M\xi^a=-S^{ab}\frac{p_b}{\mathcal M},
\end{equation}
are given in a general orthonormal frame by
\begin{align}\label{sxicomps}
\mathcal M s^0&=p_iS_i,\qquad\quad
\mathcal M s^i = p^0S_i+\epsilon_{ijk}p_j S_{0k},
\nonumber\\
\mathcal M^2\xi^0&  =p_iS_{0i},\qquad
\mathcal M^2 \xi^i =p^0S_{0i}-\epsilon_{ijk}p_jS_k,
\end{align}
with  $S_i=\tfrac{1}{2}\epsilon_{ijk}S_{jk}$ and with $S_{0i}$ determined by the SSC, as in (\ref{NWSSCSi0}) or (\ref{covSSCSi0}).

The only further ingredients needed for an explicit expression of the curvature couplings are the components of the 2-forms $\tau_{ab}$ and $\chi_{ab}$ in a given orthonormal frame.  These are given by (\ref{tauchi}) above in the curvature-aligned frame $e_a{}^\mu$, and by (\ref{tauchitaf}) below in a new time-aligned frame $f_a{}^\mu$, which we now describe.

\subsubsection{In the time-aligned frame with the Newton-Wigner SSC}

Consider the tetrad $f_a{}^\mu$ which is obtained by boosting the curvature-aligned tetrad $e_a{}^\mu$ of (\ref{CarterTetrad}) to achieve the time-aligning conditions $f^0{}_\mu=N\delta^t_\mu$ and $f_a{}^t=\delta_a^0/N$, given by
\begin{equation}\label{boost}
f_a{}^\mu=\lambda_a{}^be_b{}^\mu,
\end{equation}
where
\begin{equation}
\left(\lambda_{a}{}^{b}\right)=\left(\begin{tabular}{cccc}
 \ensuremath{\gamma} &  0  &  0  & \ensuremath{-v\gamma}\\
0  &  \ensuremath{1} &  0  &  0 \\
0  &  0  &  \ensuremath{1} &  0 \\
\ensuremath{-\gamma v} &  0  &  0  &  \ensuremath{\gamma}
\end{tabular}\right)
\end{equation}
with
\begin{equation}\label{eq:v3}
v=\frac{a\sqrt{\Delta}\sin\theta}{\varpi^2},\qquad\gamma=\frac{1}{\sqrt{1-v^2}}=\frac{\varpi^2}{\sqrt{\Lambda}},
\end{equation}
resulting in
\begin{equation}\label{SpheroidalTetrad}
\left(f_{a}{}^{\mu}\right)=\left(\begin{tabular}{cccc}
 \ensuremath{\sqrt{\dfrac{\Lambda}{\Delta\Sigma}}} &  0  &  0  &  \ensuremath{\dfrac{2Mar}{\sqrt{\Delta\Sigma\Lambda}}}\\
0  &  \ensuremath{\sqrt{\dfrac{\Delta}{\Sigma}}} &  0  &  0 \\
0  &  0  &  \ensuremath{\dfrac{1}{\sqrt{\Sigma}}} &  0 \\
0 &  0  &  0  &  \ensuremath{\dfrac{\sqrt{\Sigma}}{\sqrt{\Lambda}\sin\theta}}
\end{tabular}\right).
\end{equation}
This coincides with the ``spheroidal'' tetrad used in \cite{Barausse:Racine:Buonanno:2009}. The components of the 2-forms $\tau_{ab}$ and $\chi_{ab}$ in this frame are then obtained from (\ref{tauchi}) and (\ref{boost}) as
\begin{equation}\label{tauchitaf}
\tau_{ab}=2\gamma(\delta^0_{[a}-v^{\phantom{0}}_{[a})\delta_{b]}^1,\qquad \chi_{ab}=\frac{1}{2}\epsilon_{ab}{}^{cd}\tau_{cd},
\end{equation}
where
\begin{equation}\label{eq:vi}
v_a=\delta_a^iv_i,\qquad v_i=v\delta^3_i.
\end{equation}

It is convenient now to introduce a 3-vector notation for the spatial frame components of vectors, as in $\vec p=(p_i)$ and $\vec S=(S_i)$ with $\vec p\cdot\vec S=p_iS_i$ and $\vec p\times\vec S=(\epsilon_{ijk}p_jS_k)$.  Defining the (radial) unit vector $\vec n=(n_i)=(\delta^1_i)$ and a vector $\vec a=(a_i)$ representing the spin of the Kerr black hole,
\begin{equation}\label{eq:na}
\vec{n}=\left(\begin{tabular}{c}
1\\
0\\
0 
\end{tabular}\right),
\qquad \vec{ a}=a\left(\begin{tabular}{c}
 \ensuremath{\cos\theta}\\
\ensuremath{-\sin\theta}\\
0 
\end{tabular}\right),
\end{equation}
the boost velocity vector $\vec v=(v_i)$ from (\ref{eq:v3}) and (\ref{eq:vi}) is given by
\begin{equation}
 \vec v=-\frac{\sqrt{\Delta}}{\varpi^2}\vec n\times\vec a.
\end{equation}
Then, from (\ref{tauchitaf}) and (\ref{sxicomps}), using the solution (\ref{NWSSCSi0}) to the NW SSC, the scalars (\ref{sxicontrs}) entering the curvature couplings (\ref{gfccs}) can be written as
\begin{align}
\tau_s&=\frac{\gamma}{\mathcal M}\left[p^0\vec n\cdot \vec{\hat S}+(\vec n\times\vec v)\cdot(\vec p\times\vec S)\right],
\nonumber\\*
\chi_s&=\frac{\gamma}{\mathcal M}\left[-\vec n\cdot\vec p\times\vec S+p^0\vec n\cdot\vec v\times\vec{\hat S}\,\right],
\nonumber\\*
\tau_\xi&=\gamma\left[p^0\vec n\cdot\vec{\xi}+(\vec n\times\vec v)\cdot(\vec p\times\vec{\xi})\right],
\nonumber\\*
\chi_\xi&=\gamma\left[-\vec n\cdot\vec p\times\vec{\xi}+p^0\vec n\cdot\vec v\times\vec{\xi}\,\right],
\end{align}
where
\begin{equation}
\vec{\hat S}= \vec{S} - \frac{\vec{p} (\vec{p} \cdot \vec{S})}{p^0(p^0 + \mathcal M)},
\qquad
\vec{\xi}  =\frac{-\vec p\times\vec S}{\mathcal M(p^0+\mathcal M)},
\end{equation}
with $\vec{\hat S}$ being an auxiliary spin vector and with $\vec\xi=(\xi^i)$ as in (\ref{sxicomps}).  
We thus have fully explicit 3-vector expressions for the curvature couplings (\ref{gfccs}), if we also note that $s_as^a=\vec S^{\,2}$ and $\xi_a\xi^a=\vec\xi^{\,2}$, and that $E$, $B$, $v$, and $\gamma$ can be expressed in terms of $M$, $r$, $a$, and $\vec n\cdot\vec a=a\cos\theta$ via (\ref{Sigma_etc}), (\ref{eib}) and (\ref{eq:v3}).

\section{Explicit canonical Hamiltonian in the time-aligned frame(s) and its post-Newtonian expansion}\label{sec:PN}

Having found useful expressions for the curvature couplings, the last major step in evaluating the Hamiltonian defined by (\ref{eq:msq}) and (\ref{muHam}) in Kerr is to evaluate the Ricci rotation coefficients for a given choice of tetrad.  We address this in Sec.~\ref{sec:ricci}, giving results for the (spherical) time-aligned frame $f_a{}^\mu$, and introducing a new (Cartesian) time-aligned frame $g_a{}^\mu$ which is obtained from a spatial rotation of $f_a{}^\mu$.  We find that the PN expansion is most easily accomplished by using the rotation coefficients of the $g$-frame expressed in the $f$-frame.  We also present the rotation coefficients of the curvature-aligned frame $e_a{}^\mu$ in Appendix \ref{app:rot}. 

 In Sec.~\ref{sec:expand}, we take the fully relativistic Hamiltonian defined by the $g$-frame (expressed in the $f$-frame) and generate its PN expansion.  We are able to recover the test-mass limits of all (fully) known PN spin couplings in the center-of-mass frame through 4PN order.  Some further results (including next-to-next-to-leading-order spin-squared couplings at 4PN and next-to-leading-order spin-cubed couplings at 4.5PN) are available upon request.

\subsection{Rotation coefficients for the spherical and Cartesian time-aligned frames}\label{sec:ricci}

Given an orthonormal frame $f_a{}^\mu=(f_0{}^\mu,f_i{}^\mu)$, its Ricci rotation coefficients,
\begin{equation}\label{omegaf}
\omega^{(f)}_{abc}=f_a{}^\mu (\nabla_\mu f_b{}^\nu)f_{c\nu},
\end{equation}
with $\omega^{(f)}_{abc}=-\omega^{(f)}_{acb}$, are conveniently encoded in the two 4$\times$3 matrices
\begin{equation}\label{omegath}
{\omega}^{(f)}_{a*i}\equiv\frac{1}{2}\epsilon_{ijk}\omega^{(f)}_{ajk},\qquad\quad \omega^{(f)}_{a0i}.
\end{equation}
For the spherical time-aligned tetrad $f_a{}^\mu$ (\ref{SpheroidalTetrad}), one finds
\begin{widetext}
\begin{equation}
\left({\omega}^{(f)}_{a*i}\right)=\frac{1}{\Sigma^{3/2}\Lambda}\left(\begin{tabular}{cccc}
 \ensuremath{-2Ma^3r\sqrt{\Delta}\cos\theta\sin^2\theta} &  \ensuremath{-Ma\sin\theta(2r^2\Sigma+\varpi^2\rho^2)}  &  0 \\
0  & \phantom{\Big|}0\phantom{\Big|} &  \ensuremath{\Lambda a^2\cos\theta\sin\theta} \\
0  &  \phantom{\Big|}0\phantom{\Big|}  &  \ensuremath{\Lambda r \sqrt{\Delta}}  \\
  \ensuremath{\cot\theta(\Delta\Sigma^2+2Mr\varpi^4)} & \phantom{0}\ensuremath{-\sqrt{\Delta}(r\Sigma^2-Ma^2\rho^2\sin\theta)}\phantom{0} & 0
\end{tabular}\right),
\end{equation}
\begin{equation}
\left({\omega}^{(f)}_{a0i}\right)=\frac{1}{\Sigma^{3/2}\Lambda}\left(\begin{tabular}{cccc}
 \ensuremath{M(\varpi^4\rho^2-4Ma^2r^3\sin^2\theta)/\sqrt{\Delta}} &  \ensuremath{-2Ma^2r\varpi^2\cos\theta\sin\theta}  &  0 \\
0  & \phantom{\Big|}0\phantom{\Big|} &  \ensuremath{-Ma\sin\theta(2r^2\Sigma+\varpi^2\rho^2)} \\
0  &  \phantom{\Big|}0\phantom{\Big|}  &  \ensuremath{2Ma^3r\sqrt{\Delta}\cos\theta\sin^2\theta}  \\
  \ensuremath{-Ma\sin\theta(2r^2\Sigma+\varpi^2\rho^2)} & \phantom{0}\ensuremath{2Ma^3r\sqrt{\Delta}\cos\theta\sin^2\theta}\phantom{0} & 0
\end{tabular}\right),
\end{equation}
\end{widetext}
where
\begin{equation}
\rho^2=r^2-a^2\cos^2\theta.
\end{equation}

The same matrices for the curvature-aligned $e$-frame are given in Appendix \ref{app:rot}.  They have the same pattern of nonzero components, but the expressions for the $e$-frame components are somewhat less lengthy than the $f$-frame results given here, allowing us to easily write the exact $e$-frame coefficients in a 3-vector notation (which we do not do for the exact $f$-frame coefficients).  The advantage of the $f$-frame over the $e$-frame comes in the post-Newtonian expansion, as we can see that several components of $\omega^{(f)}_{abc}$ are shifted to higher orders in $M/r$ and $a/r$ relative to those in $\omega^{(e)}_{abc}$.

Further such simplifications for the PN expansion can be achieved by using a third ``Cartesian time-aligned'' tetrad $g_a{}^\mu$, which coincides with the ``quasi-isotropic'' tetrad of \cite{Barausse:2009xi}.  It is obtained from $f_a{}^\mu$ by a spatial rotation---the rotation that takes the spherical-coordinate triad $(e_r,e_\theta,e_\phi)$ into the Cartesian triad $(e_x,e_y,e_z)$ in flat space---,
\begin{equation}
g_a{}^\mu=\mathcal R_a{}^b f_b{}^\mu,
\end{equation}
with
\begin{equation}
\left(\mathcal R_a{}^b\right)=\left(\begin{tabular}{cccc}
1 &  0  & 0 & 0 \\
0  & \ensuremath{\sin\theta\cos\phi} &  \ensuremath{\cos\theta\cos\phi} & \ensuremath{-\sin\phi} \\
0  &  \ensuremath{\sin\theta\sin\phi}  &  \ensuremath{\cos\theta\sin\phi} & \ensuremath{\cos\phi}  \\
 0 & \ensuremath{\cos\theta} & \ensuremath{-\sin\theta} & 0
\end{tabular}\right).
\end{equation}
We will find it most convenient to use the rotation coefficients $\omega^{(g)}_{abc}$ of the $g$-frame, but expressed in the $f$-frame:
\begin{eqnarray}
\omega_{abc}&\equiv&\mathcal R^d{}_a\mathcal R^e{}_b\mathcal R^f{}_c\,\omega^{(g)}_{def}
\nonumber\\*
&=&\omega^{(f)}_{abc}-f_a{}^\mu\mathcal R_{dc}\nabla_\mu\mathcal R^d{}_b
\end{eqnarray}
where the second line follows from $f_a{}^\mu=\mathcal R^b{}_ag_b{}^\mu$, (\ref{omegaf}), and (\ref{omegaf}) with $f\to g$.  We find that the components of our hybrid rotation coefficients $\omega_{abc}$ are given by
\begin{align}\label{omegastar}
\omega_{a*i}&=\frac{1}{2}\epsilon_{ijk}\omega_{ajk}=\omega^{(f)}_{a*i}+\Delta\omega_{a*i},
\\*
\omega_{a0i}&=\omega^{(f)}_{a0i},
\end{align}
where
\begin{align}
\left(\Delta\omega_{a*i}\right)&=\left(-\frac{1}{2}\epsilon_{ijk}f_a{}^\mu\mathcal R_{bk}\nabla_\mu\mathcal R^b{}_j\right)
\\*\nonumber
&=\left(\begin{tabular}{cccc}
 \ensuremath{-\dfrac{2Mar\cos\theta}{\sqrt{\Delta\Sigma\Lambda}}} &  \ensuremath{\dfrac{2Mar\sin\theta}{\sqrt{\Delta\Sigma\Lambda}}}  &  0 \\
0  & 0 &  \phantom{\bigg|}0\phantom{\bigg|} \\
0  &  0  &  \ensuremath{-\dfrac{1}{\sqrt{\Sigma}}}  \\
  \ensuremath{-\cot\theta\sqrt{\dfrac{\Sigma}{\Lambda}}} & \ensuremath{\sqrt{\dfrac{\Sigma}{\Lambda}}} & 0
\end{tabular}\right).
\end{align}

\subsection{Post-Newtonian expansion of the Hamiltonian}\label{sec:expand}

The above results for the rotation coefficients $\omega_{abc}$ complete the list of ingredients needed for an explicit expression of the canonical Hamiltonian defined by (\ref{eq:msq}) and (\ref{muHam}).  We recall that the Hamiltonian is given by $H=-P_t$, where $P_\mu=(P_t,P_\mathrm{i})$ are the coordinate-basis components of the canonical momentum $P_a$, which is related to the covariant momentum $p_a$ by
\begin{equation}\label{Pph}
P_a-p_a=\frac{1}{2}\omega_{abc}S^{bc}\equiv h_a.
\end{equation}
The Hamiltonian $H=-P_t$ is found by solving the mass shell constraint (\ref{eq:msq}), $p^2=-\mathcal M^2$ $\Rightarrow$
\begin{align}\label{muh}
\mu^{2}&=-P^2 
\\
&= m^{2}-2P^ah_a+h^ah_a
\nonumber\\\nonumber
&\quad+\frac{1}{4}R_{abcd}S^{ab}S^{cd}-(C-1)E^{(p)}_{ab}s^as^b+\mathcal{O}(S^{3}),
\end{align}
for $P_t$.  A formal solution is given by (\ref{muHam}) above, and the solution explicitly expanded to quadratic order in the test spin is given by (\ref{quadHamil}) below.  

The following subsections collect results for the Kerr-spin and PN expansions of the rotation coefficients, the spatial triad, and the metric coefficients, and for the test-spin expansion of the Hamiltonian.  The results of the PN expansion are then presented and discussed in Sec.~\ref{sec:expanded}.

\subsubsection{Expansion of the rotation coefficients}\label{sec:expandricci}

The components of the rotation coefficients are conveniently encoded in the vector $h_a$ of (\ref{Pph}), which is expressed via (\ref{omegastar}) as
\begin{equation}\label{hstar}
h_a=\frac{1}{2}\omega_{abc}S^{bc}=\omega_{a*i}S_i-\omega_{a0i} S_{0i}.
\end{equation}
The results can be given in a 3-vector notation by expressing the components $h_0$ and $\vec h=(h_i)$ in terms of $\vec S=(S_i)$ and $\vec S_0=(S_{0i})$. The only other 3-vectors that will appear in these expressions are the radial unit vector $\vec n$ and the Kerr spin vector $\vec a$ of (\ref{eq:na}), and the only other quantities involved are the Kerr mass $M$ and the Boyer-Lindquist radial coordinate $r$.
We present the results here as an expansion in the Kerr spin $a$:
\begin{align}
h_0&=h_0^{a^0}+h_0^{a^1}+h_0^{a^2}+h_0^{a^3}+\mathcal O(a^4),
\\
\vec h&=\vec h^{a^0}+\vec h^{a^1}+\vec h^{a^2}+\vec h^{a^3}+\mathcal O(a^4).
\end{align}
At $\mathcal O(a^0)$ and $\mathcal O(a^1)$, keeping all powers of $M/r$, we have
\begin{align}
h_0^{a^0}&=-\frac{M}{r^2\sqrt{w}}\vec n\cdot\vec S_0,
\\
\vec h^{a^0}&=\frac{1-\sqrt{w}}{r}\vec n\times\vec S,
\\
h_0^{a^1}&=\frac{M}{r^3}\left[-3\vec n\cdot\vec a\;\vec n\cdot\vec S+\left(3-\frac{2}{\sqrt{w}}\right)\vec a\cdot\vec S\right],
\\
\vec h^{a^1}&=-\frac{3M}{r^3}\left[\vec n\cdot\vec S_0\;\vec n\times\vec a+(\vec n\cdot\vec a\times\vec S_0)\,\vec n\right],
\end{align}
where
\begin{equation}\label{eq:w}
w=1-\frac{2M}{r}.
\end{equation}
At $\mathcal O(a^2)$ and $\mathcal O(a^3)$, expanding in $M/r$, we have
\begin{align}
h_0^{a^2}&=h_0^{\mathrm{LO}a^2}+h_0^{\mathrm{NLO}a^2}+\mathcal O\left(\frac{M^3a^2S_0}{r^6}\right),
\\\nonumber
\vec h^{a^2}&=\vec h^{\mathrm{PLO}a^2}+\vec h^{\mathrm{LO}a^2}+\vec h^{\mathrm{NLO}a^2}+\mathcal O\left(\frac{M^3a^2S}{r^6}\right),
\end{align}
with
\begin{align}
\vec h^{\mathrm{PLO}a^2}&=\frac{1}{2r^3}\left[\vec n\cdot\vec a\;\vec a\times\vec S+\vec n\cdot\vec a\times\vec S\Big(\vec a-4\vec n\cdot \vec a\;\vec n\Big)\right],
\nonumber\\
h_0^{\mathrm{LO}a^2}&=\frac{M}{2r^4}\Big[-4\vec n\cdot\vec a\;\vec a\cdot\vec S_0
\\\nonumber
&\;\;\quad\qquad+\Big(\!-\!\vec a^{\,2}+11(\vec n\cdot\vec a)^2\Big)\vec n\cdot\vec S_0\Big],
\\\nonumber
\vec h^{\mathrm{LO}a^2}&=\frac{M}{2r^4}\Big[
-\Big(\vec a^{\,2}+3(\vec n\cdot\vec a)^2\Big)\vec n\times\vec S
\\\nonumber
&\qquad\quad\;\;+2\Big(\vec a\cdot\vec S-4\vec n\cdot\vec a\;\vec n\cdot\vec S\Big)\vec n\times\vec a\Big],
\\\nonumber
h_0^{\mathrm{NLO}a^2}&=\frac{M^2}{2r^5}\Big(13\vec a^{\,2}-5(\vec n\cdot\vec a)^2\Big)\vec n\cdot\vec S_0,
\\\nonumber
\vec h^{\mathrm{NLO}a^2}&=\frac{M^2}{4r^5}\Big[
-3\Big(\vec a^{\,2}+(\vec n\cdot\vec a)^2\Big)\vec n\times\vec S
\\\nonumber
&\qquad\quad\;\;+14\Big(-\vec a\cdot\vec S+\vec n\cdot\vec a\;\vec n\cdot\vec S\Big)\vec n\times\vec a\Big],
\end{align}
and
\begin{align}
h_0^{a^3}&=h_0^{\mathrm{LO}a^3}
+\mathcal O\left(\frac{M^2a^3S}{r^6}\right),
\\*\nonumber
\vec h^{a^3}&=\mathcal O\left(\frac{Ma^3 S_0}{r^5}\right),
\end{align}
with
\begin{equation}
h_0^{\mathrm{LO}a^3}=\frac{M}{2r^5}(\vec n\cdot\vec a)^2\Big(17 \vec n\cdot\vec a\;\vec n\cdot\vec S-9\vec a\cdot\vec S\Big).
\end{equation}

\subsubsection{Expansion of the spatial triad; coordinate-basis versus frame components of the canonical momentum}

While our expressions for the spin coupling terms in (\ref{muh}) involve the spatial components $P_i=P^i$ of the $f$-frame components $P^a=(P^0,P^i)$ of the canonical momentum, our true canonical variables are the spatial components $P_{\,\mathrm{i}}$ of the coordinate-basis components $P_\mu=(P_t,P_{\,\mathrm{i}})$ of the canonical momentum.  The two are related by $P_i=f_i{}^\mathrm{j}P_\mathrm{j}$, where $f_i{}^\mathrm{j}=\mathrm{diag}\left(\sqrt{\Delta/\Sigma},1/\sqrt{\Sigma},\sqrt{\Sigma}/(\sqrt{\Lambda}\sin\theta)\right)$, from (\ref{SpheroidalTetrad}).  Writing these two sets of components in 3-vector notation, with $(e_\mathrm{flat})_i{}^{\mathrm{i}}=\big(\mathrm{diag}(1,1/r,1/r\sin\theta)\big)_i{}^{\mathrm{i}}$,
\begin{align}
\vec P_{(f)}=(P_i),
\qquad
\vec P=\big((e_\mathrm{flat})_i{}^{\mathrm{i}}P_{\,\mathrm{i}}\big)
\end{align}
(which are in two distinct spherical-like orthonormal bases [$(e_r, e_\theta, e_\phi)$] which will now be identified component-wise),
the explicit relationship $P_i=f_i{}^\mathrm{j}P_\mathrm{j}$, expanded in the Kerr spin but not in PN orders, is given by
\begin{align}
\vec P_{(f)}&=\vec{P}-(1-\sqrt{w})\vec n\cdot\vec P\;\vec n
\\*
&\quad+\frac{1}{2r^2}\bigg[\left(\frac{\vec a^{\,2}}{\sqrt{w}}+(1-\sqrt{w})(\vec n\cdot\vec a)^2\right)\vec n\cdot\vec P\;\vec n
\nonumber\\*\nonumber
&\quad-(\vec n\cdot\vec a)^2\vec P+(2-w)(\vec n\cdot\vec P\times\vec a)\, \vec n\times\vec a\bigg]+\mathcal O(a^4).
\end{align}
The translation needed to connect with the curvature coupling results of Sec.~\ref{sec:NWEB},
where we denoted the spatial $f$-frame components of the covariant momentum as the 3-vector $\vec p=(p_i)$, is simply $\vec p=\vec P_{(f)}+\mathcal O(S)$, from (\ref{Pph}).

\subsubsection{Expansion of the metric coefficients}\label{sec:lapse}
The lapse $N$, shift $N^\mathrm{i}$, and inverse spatial metric $\gamma^\mathrm{ij}$ of (\ref{lapseshift}) are given by
\begin{align}
N=\sqrt{\frac{\Delta\Sigma}{\Lambda}}
&=\sqrt{w}+\frac{M}{r^3}\left(\sqrt{w}(\vec n\cdot\vec a)^2+\frac{2M\vec a^{\,2}}{r\sqrt{w}}\right)
\\*\nonumber
&\quad-\frac{M}{r^5}(\vec n\cdot\vec a)^4+\mathcal O\left(\frac{M^2a^4}{r^6}\right)+\mathcal O(a^6),
\end{align}
\begin{align}
\vec N&=\big((e_\mathrm{flat})^i{}_{\mathrm{i}}N^{\mathrm{i}}\big)=-\frac{2Mar^2\sin\theta}{\Lambda}(\delta^i_3)
\\*\nonumber
&=\frac{2M}{r^2}\vec n\times\vec a\left(1-\frac{w(\vec n\cdot\vec a)^2+(2-w)\vec a^{\,2}}{r^2}\right)+\mathcal O(a^5),
\end{align}
and
\begin{align}
&\gamma^\mathrm{ij}P_{\,\mathrm{i}}P_{\,\mathrm{j}}=P_iP_i=\vec P_{(f)}\cdot \vec P_{(f)}
\\*
&=\vec P\cdot \mathrm{diag}\left({\Delta/\Sigma},r^2/{\Sigma},{r^2\Sigma}/{\Lambda}\right)\cdot\vec P
\nonumber\\*\nonumber
&=\vec P^{\,2}-\frac{2M}{r}(\vec n\cdot\vec P)^2
\\*
&\quad+\frac{1}{r^2}\Big(\vec a^{\,2}(\vec n\cdot\vec P)^2-(\vec n\cdot\vec a)^2\vec P^{\,2}-(\vec n\cdot\vec P\times\vec a)^2\Big)
\nonumber\\*
&\quad+\frac{2M}{r^3}\Big((\vec n\cdot\vec a)^2(\vec n\cdot\vec P)^2-(\vec n\cdot\vec P\times\vec a)^2\Big)
\nonumber\\*
&\quad+\frac{1}{r^4}\bigg[(\vec n\cdot\vec a)^2\Big((\vec n\cdot\vec a)^2\vec P^{\,2}-\vec a^{\,2}(\vec n\cdot\vec P)^2\Big)
\nonumber\\*\nonumber
&\quad+\Big(\vec a^{\,2}+(\vec n\cdot\vec a)^2\Big)(\vec n\cdot\vec P\times\vec a)^2\bigg]+\mathcal O\left(\frac{Ma^4}{r^5}\right)+\mathcal O(a^6).
\end{align}

\subsubsection{Expansion in powers of the test spin}\label{sec:expandinspin}

The explicit expression of the Hamiltonian $H=-P_t$, expanded to quadratic order in the test spin---in a general spacetime, using the frame components of a general tetrad obeying the time-aligning conditions---can be found by perturbatively solving equation (\ref{muh}) for $P_t$, using (\ref{Pph}), (\ref{S0s}), and (\ref{lapseshift}).  One finds

\begin{align}\label{quadHamil}
H&=NQ-N^{\mathrm{i}}P_{\,\mathrm{i}}+\frac{N}{Q}\Bigg[-\hat P^a\hat h_a+\frac{\hat h^a \hat h_a}{2}-\frac{(\hat P^a \hat h_a)^2}{2Q^2}
\nonumber\\
&\quad+\frac{\hat h_0}{Q}\hat P^a\hat h_a-\frac{\hat P^a\omega_{a0i}S^{ij}}{Q+m}\left(\hat  h_{j}-\frac{P_{j}P^{k}\hat h_{k}}{Q(Q+m)}\right)
\nonumber\\
&\quad+\frac{1}{8}R_{abcd}S^{ab}S^{cd}-\frac{C-1}{2}E^{(P)}_{ab}s^as^b\Bigg]+\mathcal O(S^3),
\end{align}
where 
\begin{equation}\label{eq:PQ}
\hat P^a=(Q,P^i), \qquad Q=\sqrt{m^2+\gamma^\mathrm{ij}P_{\,\mathrm{i}}P_{\,\mathrm{j}}},
\end{equation}
and where
\begin{equation}\label{EQ:h}
\hat h_a=\omega_{a*i}S_i-\omega_{a0i}\hat S_{0i},\qquad \hat S_{0i}=\frac{\epsilon_{ijk}P_jS_k}{Q+m}
\end{equation}
are the solutions for $h_a$ and $S_{0i}$ to linear order in the test spin.

\subsection{Results of the post-Newtonian expansion}\label{sec:expanded}

Taking the results of Sec.~\ref{sec:NWEB} for the curvature couplings in the $f$-frame, and the results of Secs.~\ref{sec:expandricci}-\ref{sec:lapse} for the expanded $f$-$g$-hybrid-frame rotation coefficients and spatial triad and the metric, and substituting them into (\ref{quadHamil}) yields the explicit expression of the Hamiltonian $H(\vec z,\vec P,\vec S)$, where $\vec z=r\vec n$, $\vec n=(\delta^1_i)$, $\vec P=\big((e_\mathrm{flat})_i{}^{\mathrm{i}}P_{\,\mathrm{i}}\big)$, and $\vec S=(S_i)$.  It is in a form which can be easily expanded in powers of the Kerr spin $a$ and then further expanded in PN orders, measured by the PN parameter
\begin{equation}
\epsilon\sim\frac{M}{r}\sim\frac{P^2}{m^2}\sim v^2.
\end{equation}
Details of this procedure are provided in accompanying Mathematica notebooks using the xTensor package \cite{xTensor}, available upon request.

We summarize the results in the following subsections, going order by order in the Kerr spin $S_\mathrm{Kerr}=Ma$ and the test spin $S$, 
\begin{align}
H&=H_\mathrm{p.p.}+H_a+H_S
\\\nonumber
&\qquad\quad\;\;+H_{a^2}+H_{aS}+H_{S^2}
\\\nonumber
&\qquad\quad\;\;+H_{a^3}+H_{a^2S}+H_{aS^2}+\left\{H_{S^3}\right\}
\\\nonumber
&\qquad\quad\;\;+H_{a^4}+H_{a^3S}+H_{a^2S^2}+\left(H_{aS^3}\right)+\left(H_{S^4}\right)
\\\nonumber
&\qquad\quad\;\;+\mathcal O(a,S)^5.
\end{align}
For the first two lines, through spin-squared order, we give explicit results for $H$ to all powers in the PN parameter $\epsilon$ in Appendix \ref{app:allPN}.  Instead of giving explicit results for our $H$ at third and fourth orders in the spin, we give PN expanded results for a Hamiltonian $\bar H$ which is obtained from a canonical transformation of our $H$ via a generating function $\mathcal G(\vec z,\vec P,\vec S)$ according to
\begin{equation}\label{cantrans}
\bar H=H+\{\mathcal G,H\}+\frac{1}{2}\left\{\mathcal G,\{\mathcal G,H\}\right\}+\ldots\,
\end{equation}
The canonical transformation brings our Hamiltonian into accord with the test-mass limits of PN results obtained in harmonic coordinates, as detailed below.

Our test-spin-squared Hamiltonian does not allow us to compute the contributions $H_{S^3}$, $H_{aS^3}$, and $H_{S^4}$.  However, one finds from the PN calculation of \cite{Levi:Steinhoff:2014:2} that, at the leading PN order (LO), for binary black holes ($C=1$),
\begin{align}\label{deduce13}
\bar H_{\mathrm{LO}aS^3}&=\bar H_{\mathrm{LO}a^3S}(M\vec a\leftrightarrow \vec S),
\\\label{deduce04}
\bar H_{\mathrm{LO}S^4}&=\bar H_{\mathrm{LO}a^4}(M\vec a\leftrightarrow \vec S),
\end{align}
so that these results as well can be ``deduced'' from our test-spin-squared Hamiltonian.  (Note the correspondence here with the EOB prescription of replacing the Kerr spin with the sum of the two individual spins \cite{Barausse:2009xi}.)  At the leading PN orders, the true (finite-mass-ratio) Hamiltonians at second and fourth orders in spin are equal to their test-body limits, so that all of these contributions \emph{for finite mass ratios} (for binary black holes) can be deduced from our Hamiltonian.  The situation for $H_{S^3}$ is different because, even at leading PN order, the spin-cubed terms (like the linear-in-spin terms) have contributions at zeroth and first orders in the mass ratio $m/M$, only one set of which can be deduced from our Hamiltonian by exchanging the bodies.  However, the complete finite-mass-ratio leading-PN-order results at first and third orders in spin (with a restriction to binary black holes for the third-order terms) can still be deduced from the test-body limit, exploiting a body-exchange symmetry (given an extension of the test-body results to cubic order in the test-spin).

\begin{figure*}
\includegraphics[scale=1]{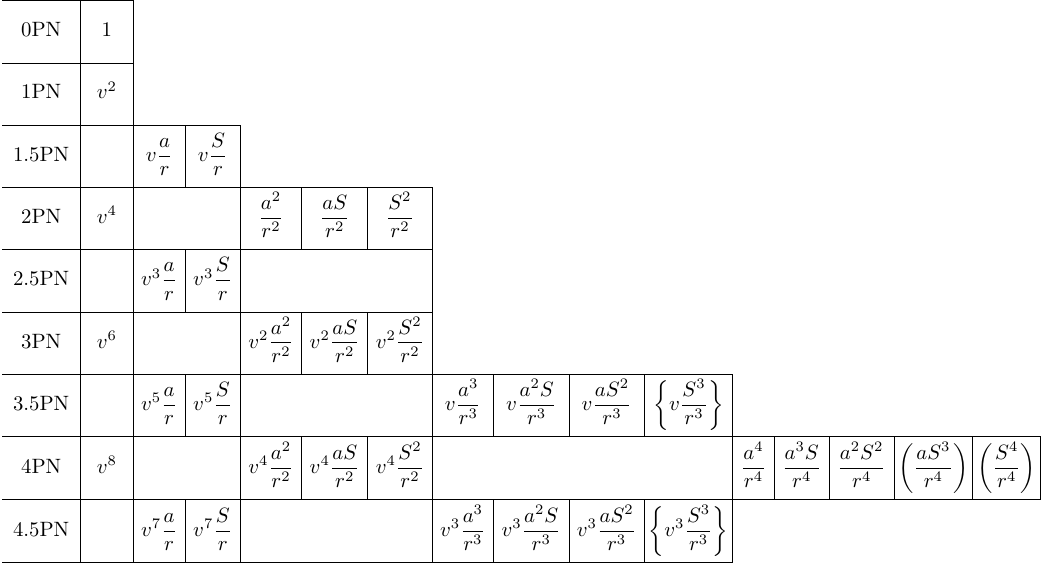}
\caption{
Orders (relative to Newtonian order) of the PN-spin expansion through 4.5PN, counting $\epsilon\sim v^2\sim M/r\sim a/r\sim S/r$ (ignoring the scalings with the mass ratio) for rapidly rotating bodies.  
}
\end{figure*}
We now present the PN-expanded results for the test-body Hamiltonian in a form in which the quantities $\vec P$, $\vec S$, $\bar H$ and $\mathcal G$ have each absorbed one factor of $1/m$, which removes all factors of $m$ from the equations (i.e., we set $m=1$).

\subsubsection{No spin}

The Hamiltonian to zeroth order in $a$ and $S$, the point-particle Hamiltonian, is given to all PN orders by (\ref{allpp}).  Its expansion through 2PN, transformed with the contribution
\begin{equation}
\mathcal G_\mathrm{1PN}=-M\vec n\cdot\vec P
\end{equation}
to the generating function in (\ref{cantrans}), with no $\mathcal G_\mathrm{2PN}$ needed, yields
\begin{align}
\bar H_\mathrm{p.p.}&=\frac{\vec P^{\,2}}{2}-\frac{M}{r}
\\*\nonumber
&\quad-\frac{\vec P\,^4}{8}-\frac{3M \vec P^{\,2}}{2r}+\frac{M^2}{2r^2}
\\*\nonumber
&\quad+\frac{\vec P\,^6}{16}+\frac{5M\vec P\,^4}{8r}+\frac{5M^2\vec P^{\,2}}{2r^2}-\frac{M^3}{4r^3}+\mathcal O(\epsilon^4),
\end{align}
which matches the test-body limit of Eqs.~(4.23-4.25) of Ref.~\cite{Levi:Steinhoff:2015:2} (and the original calculation in this gauge of Eq.~(63) of Ref.~\cite{Gilmore:Ross:2008}, after a Legendre transformation).  The 0PN terms in the first line get no mass-ratio corrections in the true Hamiltonian, the 1PN terms in the second line get corrections at linear order in the mass ratio, the 2PN terms in the third line get corrections through second order in the mass ratio, and so on.

\subsubsection{Linear in spin}

The Hamiltonians at linear order in the spins are given to all PN orders by (\ref{alla}) and (\ref{allS}).  Their expansions through next-to-leading order (NLO) and next-to-next-to-leading order (NNLO), subjected to the canonical transformation (\ref{cantrans}) with
\begin{equation}
\mathcal G_{\mathrm{NNLO}S}=-\frac{3M^2}{r^2}\vec n\cdot\vec P\;\vec n\cdot\vec P\times\vec S
\end{equation}
as the only new contribution (being added to $\mathcal G_\mathrm{1PN}$), yield
\begin{equation}
\bar H_{a}=\frac{M}{r^2} \vec n\cdot\vec P\times \vec a\left[2-\frac{6M}{r}+\frac{12M^2}{r^2}+\mathcal O(\epsilon^3)\right]
\end{equation}
and
\begin{align}
&\bar H_{S}=\frac{M}{r^2} \vec n\cdot\vec P\times \vec S\bigg[\frac{3}{2}
-\frac{5\vec P^{\,2}}{8}-\frac{5M}{r}
\\*\nonumber
&+\frac{7\vec P\,^4}{16}+\frac{21M\vec P^{\,2}}{8r}+\frac{3M(\vec n\cdot\vec P)^2}{r}+\frac{45M^2}{4r^2}+\mathcal O(\epsilon^3)\bigg] ,
\end{align}
which match (e.g.)\ the test-body limits of Eqs.~(4.26-4.28) of Ref.~\cite{Levi:Steinhoff:2015:2}.  The LO terms get corrections at linear order in the mass ratio which can be deduced from the test-body results via exchange of the bodies.  The NLO terms get corrections through second order in the mass ratio, the NNLO terms get contributions through third order in the mass ratio, and so on.

\subsubsection{Quadratic in spin}

The Hamiltonians at quadratic order in the spins are given to all PN orders by (\ref{allaa}), (\ref{allaS}), and (\ref{allSS}).  At the leading PN order, with
\begin{equation}
\mathcal G_{\mathrm{LO}a^2}=\frac{1}{2r}(\vec n\times\vec a)\cdot(\vec P\times\vec a),
\end{equation}
and no other $\mathcal G$'s, we have
\begin{align}
\bar H_{\mathrm{LO}a^2}
&=\frac{M}{2r^3}\Big(3(\vec n\cdot\vec a)^2-\vec a^{\,2}\Big),
\\
\bar H_{\mathrm{LO}aS}&=\frac{M}{r^3}\left(3\,\vec n\cdot\vec a\;\vec n\cdot\vec S-\vec a\cdot\vec S\right),
\\
\bar H_{\mathrm{LO}S^2}&=\frac{CM}{2r^3}\Big(3(\vec n\cdot\vec S)^2-\vec S^{\,2}\Big),
\end{align}
which match Eqs.~(4.29, 31) of Ref.~\cite{Levi:Steinhoff:2015:2} with no mass ratio corrections.

At the next-to-leading PN order, with
\begin{align}
\mathcal G_{\mathrm{NLO}a^2}&=\frac{M}{4r^2}\left[\vec n\cdot\vec P\,\Big(\vec a^{\,2}+(\vec n\cdot\vec a)^2\Big)-2\vec n\cdot\vec a\;\vec P\cdot\vec a\right],\phantom{\Bigg|}
\nonumber\\\nonumber
\mathcal G_{\mathrm{NLO}aS}&=\frac{M}{r^2}\Big(\vec P\cdot\vec a\;\vec n\cdot\vec S-\vec n\cdot\vec P\;\vec a\cdot\vec S\Big),
\\
\mathcal G_{\mathrm{NLO}S^2}&=-\frac{CM}{r^2}\vec n\cdot\vec S\;\vec P\cdot\vec S,
\end{align}
we find
\begin{widetext}
\begin{align}
\bar H_{\mathrm{NLO}a^2}&=\frac{M}{r^3}\left[\left(-\frac{3\vec P^{\,2}}{4}+\frac{5M}{2r}\right)\vec a^{\,2}
+\left(\frac{9\vec P^{\,2}}{4}-\frac{9M}{2r}\right)(\vec n\cdot\vec a)^2\right],
\\
\bar H_{\mathrm{NLO}aS}&=\frac{M}{r^3}\Bigg[\left(-\frac{5\vec P^{\,2}}{2}+6(\vec n\cdot\vec P)^2+\frac{7M}{r}\right)\vec a\cdot\vec S
+\left(\frac{3\vec P^{\,2}}{2}-\frac{13M}{r}\right)\vec n\cdot\vec a\;\vec n\cdot\vec S
\\\nonumber
&\qquad\quad+\frac{5}{2}\vec P\cdot\vec a\;\vec P\cdot\vec S-\frac{3}{2}\vec n\cdot\vec P\,\Big(\vec n\cdot\vec a\;\vec P\cdot\vec S+4\vec P\cdot\vec a\;\vec n\cdot\vec S\Big)\Bigg],
\\
\bar H_{\mathrm{NLO}S^2}&=\frac{M}{r^3}\Bigg[\left(\frac{5}{4}(1-C)\vec P^{\,2}+\frac{3}{8}(4C-3)(\vec n\cdot\vec P)^2+(2C+1)\frac{M}{r}\right)\vec S^{\,2}
\\*\nonumber
&\qquad\quad+\left(\frac{3}{8}(6C-7)\vec P^{\,2}-(1+5C)\frac{M}{r}\right)(\vec n\cdot\vec S)^2
+\frac{1}{4}(2C-5)(\vec P\cdot\vec S)^2+\frac{3}{4}(5-2C)\vec n\cdot\vec P\;\vec n\cdot\vec S\;\vec P\cdot\vec S\Bigg],
\end{align}
\end{widetext}
which match the test-body limits of Eqs.~(4.30) and (4.32) of Ref.~\cite{Levi:Steinhoff:2015:2}.

At next-to-next-to-leading order, results for $H_{\mathrm{NNLO}aS}$ are available in Ref.~\cite{Levi:Steinhoff:2014:1}, and we have found agreement with the test-body limits of those results.  These were based on simultaneous calculations using different methods \cite{Hartung:2011ea, Levi:2011eq}.  However, those results are presented in a different gauge from the other results presented here.  Details are available upon request.  Results for $H_{\mathrm{NNLO}a^2}$ and $H_{\mathrm{NNLO}S^2}$ are available in principle from Ref.~\cite{Levi:Steinhoff:2015:3}, but those results are not yet sufficiently reduced to allow a comparison.  The expressions of our untransformed NNLO-spin-squared Hamiltonians are available upon request.

\subsubsection{Cubic in spin}
At third order in the spins, at leading PN order, with
\begin{align}
\mathcal G_{\mathrm{PLO}a^2S}&=\frac{1}{2r^2}\vec n\cdot\vec a\;\vec n\cdot\vec a\times\vec S,
\\
\mathcal G_{\mathrm{LO}a^2S}&=\frac{M}{2r^3}\vec n\cdot\vec a\;\vec n\cdot\vec a\times\vec S,
\\
\mathcal G_{\mathrm{LO}aS^2}&=-\frac{3CM}{r^3}\vec n\cdot\vec S\;\vec n\cdot\vec a\times\vec S,
\end{align}
we find
\begin{widetext}
\begin{align}
\bar H_{\mathrm{LO}a^3}&=\frac{M}{r^4}\Big(\vec a^{\,2}-5(\vec n\cdot\vec a)^2\Big)\vec n\cdot\vec P\times\vec a,
\\
\bar H_{\mathrm{LO}a^2S}
&=\frac{9M}{4r^4}\left[\Big(\vec a^{\,2}-5(\vec n\cdot\vec a)^2\right)\vec n\cdot\vec P\times\vec S-2\,\vec n\cdot\vec a\;\vec P\cdot\vec a\times\vec S\Big],
\\
\bar H_{\mathrm{LO}aS^2}
&=\frac{3CM}{r^4}\left[\left(\vec S^{\,2}-5(\vec n\cdot\vec S)^2\right)\vec n\cdot\vec P\times\vec a+2\,\vec n\cdot\vec S\;\vec P\cdot\vec a\times\vec S\right],
\\\nonumber
&\quad+\frac{3M}{r^4}\left[\left(5\,\vec n\cdot\vec a\;\vec n\cdot\vec S-\vec a\cdot\vec S\right)\vec n\cdot\vec P\times\vec S+\vec n\cdot\vec S\;\vec P\cdot\vec a\times\vec S\right],
\end{align}
\end{widetext}
which match the test-body limit of Eq.~(3.10) of Ref.~\cite{Levi:Steinhoff:2014:2}.  These, together with $H_{\mathrm{LO}S^3}$, receive corrections at linear order in the mass ratio.  For binary black holes, all but one term of these can be deduced from our Hamiltonian via body exchanges, with the final term requiring a treatment of test-spin-cubed effects to be derived from the test-body limit.  Expressions for our untransformed NLO-spin-cubed Hamiltonians are available upon request.

\subsubsection{Quartic in spin}

At fourth order in the spins, at leading PN order, with
\begin{equation}
\mathcal G_{\mathrm{LO}a^4}=-\frac{M}{4r^3}(\vec n\cdot\vec a)^2\,(\vec n\times\vec a)\cdot(\vec P\times\vec a),
\end{equation}
and no other $\mathcal G$'s, we find
\begin{widetext}
\begin{align}
\bar H_{\mathrm{LO}a^4}&=\frac{M}{8r^5}\left[-3\,\vec a^{\,4}+30\,\vec a^{\,2}(\vec n\cdot\vec a)^2-35(\vec n\cdot\vec a)^4\right],
\\*
\bar H_{\mathrm{LO}a^3S}&=\frac{M}{2r^5}\left[-3\,\vec a^{\,2}\,\vec a\cdot\vec S +15(\vec n\cdot\vec a)^2\,\vec a\cdot\vec S+15\,\vec a^{\,2}\,\vec n\cdot\vec a\;\vec n\cdot\vec S-35 (\vec n\cdot\vec a)^3\,\vec n\cdot\vec S\right],
\\*
\bar H_{\mathrm{LO}a^2S^2}&=\frac{3CM}{4r^5}\left[-\vec a^{\,2}\vec S^{\,2}-2(\vec a\cdot\vec S)^2+5(\vec n\cdot\vec a)^2\vec S^{\,2}+5\,\vec a^{\,2}(\vec n\cdot\vec S)^2+20\,\vec n\cdot\vec a\;\vec n\cdot\vec S\;\vec a\cdot\vec S-35(\vec n\cdot\vec a)^2(\vec n\cdot\vec S)^2
\right],
\end{align}
\end{widetext}
which match the test-body limit of Eq.~(4.4) of Ref.~\cite{Levi:Steinhoff:2014:2}.  These, along with $H_{\mathrm{LO}aS^3}$ and $H_{\mathrm{LO}S^4}$, receive no mass ratio corrections, and the latter can be deduced from our Hamiltonian via (\ref{deduce13}) and (\ref{deduce04}) for binary black holes.

\section{Discussion}\label{sec:discussion}

In this paper, we derived  a canonical Hamiltonian for an extended test body in a curved background which is valid to quadrupolar order in the multipole expansion and includes spin-induced quadrupoles as well as all other spin-squared effects. We employed a new approach that avoids the Dirac brackets used in previous work and instead enables working with an arbitrary spin supplementary condition at the level of a constrained action principle. This method provides substantial simplifications of previous calculations at the dipolar order and yields novel results at quadrupolar order.  We highlighted how a change of the SSC, corresponding to a shift of the center-of-mass worldline, which we treated in a manifestly covariant manner using bitensor calculus, entails transformations of the body's multipole moments, and exhibited the resulting modifications of the action. While our analysis focused primarily on variables determined by the Newton-Wigner SSC, we 
provided the explicit translation into variables defined by other SSCs. 

We constructed a general Hamiltonian in terms of three-dimensional position, momentum, and spin variables with a canonical Poisson bracket structure to quadratic order in the spin, given in Eqs.~(\ref{introH}) and (\ref{intromu}), or (\ref{eq:msq}) and (\ref{muHam}).  By specializing the above general results to the case where the background spacetime is Kerr, we arrived at an explicit expression for the canonical Hamiltonian of the relativistic spinning two-body problem in the test-body limit, valid to quadrupolar order in the test body's multipole expansion, given in Eq.~(\ref{quadHamil}) for a general choice of tetrad whose timelike vector is along the direction of the time coordinate.

 Our results for the dynamics allow for fully generic orbits and spin orientations, both of which have not been considered before. We provided compact expressions for curvature couplings valid for generic orbits in Eqs.~(\ref{gfccs}-\ref{sxicomps}) in a general frame and provided explicit results for two different choices of frame. 
 
 Expanding the Hamiltonian in powers of the Kerr spin and in PN orders allowed us to make comparisons with the test-body limits of the results of high-order PN calculations.  We found complete agreement with the test-body limits of all available PN results and can obtain new test-body results at higher PN orders.  We also pointed out how the complete finite-mass-ratio PN results for the leading-PN-order spin couplings for binary black holes, through fourth order in the spins, can all be inferred from the results in the test-mass limit through an EOB-like identification of variables.

While much of our analysis and many of our intermediate results are fully covariant, our final result for the Hamiltonian in Kerr depends on a choice of coordinates and a choice of tetrad.  We showed that expansion of the Hamiltonian and comparison with PN results can be relatively easily accomplished using Boyer-Lindquist coordinates and the ``quasi-isotropic'' tetrad, though we also explicitly related this tetrad to the tetrad used by Carter which diagonalizes the electric and magnetic components of the Weyl tensor.  Other choices of coordinates and tetrads are likely to yield other useful forms of the Hamiltonian and can readily be used in the general expressions we provide for the Hamiltonian.

\acknowledgements
We thank Abraham Harte and Alessandra Buonanno for stimulating and encouraging conversations.  
\mbox{D.\ K.{}} gratefully acknowledges support from the Deutsche
Forschungsgemeinschaft within the Research Training
Group 1620 ``Models of Gravity'' and from the ``Centre
for Quantum Engineering and Space-Time Research
(QUEST)'', and T.\ H.\ gratefully acknowledges support from NSF Grant No. PHY-1208881.  D.\ K.\ and T.\ H.\ also thank the Max-Planck-Institut f\"ur Gravitationsphysik for hospitality.

\section*{Note Added}

We are grateful to Dimitrios Kosmopoulos and  Andr\'es Luna for pointing us toward errors in the previously published versions of Eqs.~(\ref{allSS}) and (\ref{kcoeffs}) giving the test-spin-squared contributions to the Hamiltonian to all PN orders (for zero Kerr spin).  These were copy errors only in Eqs.~(\ref{allSS}) and (\ref{kcoeffs}) which did not affect the comparisons to PN results in Sec.~\ref{sec:expanded}.  We have checked that the new expressions (\ref{allSS}) and (\ref{kcoeffs}) in this version, along with (\ref{allpp})--(\ref{allaS}), expanded to the appropriate orders, are in complete agreement (modulo a canonical transformation) with the extended test-body limits of both the second post-Minkowskian spin-sqaured results of Kosmopoulos and Luna \cite{Kosmopoulos:2021zoq} and the next-to-next-to-leading-order post-Newtonian (4PN) spin-squared results of Levi and Steinhoff \cite{Levi:2016ofk}.

\appendix

\section{Variation of the action}\label{sec:vary}
Here we show that the variation of the action (\ref{eq:Gaction}) for a generic SSC,
$\mathcal S=\int ds \;L$, with
\begin{align}\label{app:action}
L&= p_{\mu} \dot {z}^{\mu}+\frac{1}{2}S_{\mu\nu}\Omega^{\mu\nu}-\chi^{\mu}S_{\mu\nu} \left(\frac{p^\nu}{\sqrt{-p^2}}+\Lambda_0{}^\nu\right)
\\
&\quad-\frac{\lambda}{2}\Big( p^2+{{\mathcal M}}^2(p,S,z)\Big),
\nonumber
\end{align}
 leads to the quadrupolar MPD equations (\ref{eq:MPDp}, \ref{eq:MPDS}).  This also shows that the covariant-SSC action (\ref{eq:Taction}) yields the MPD equations, as (\ref{eq:Taction}) is a special case of (\ref{eq:Gaction}), when the gauge field $\Lambda_0{}^\mu$ is taken to be $p^\mu/\sqrt{-p^2}$.  See also \cite{Marsat:2014xea} for a related derivation.

The independent variables to be varied are $p_\mu$, $S_{\mu\nu}$, $z^\mu$, and $\Lambda_A{}^\mu$ (along with the Lagrange multipliers $\lambda$ and $\chi^\mu$).
The variations with respect to $p$ and $S$ are straightforward.  Under $p_\mu\to p_\mu+\delta p_\mu$ and $S_{\mu\nu}\to S_{\mu\nu}+\delta S_{\mu\nu}$, the linear variations of the Lagrangian are
\begin{align}\label{dpL}
\delta_pL&=\left[\dot z^\mu-\lambda p^\mu-\chi^\alpha S_{\alpha\nu}\frac{\mathcal P^{\mu\nu}}{\sqrt{-p^2}}-\frac{\lambda}{2}\frac{\partial\mathcal M^2}{\partial p_\mu}\right]\delta p_\mu,
\\\nonumber
\delta_SL&=\frac{1}{2}\left[\Omega^{\mu\nu}-2\chi^{[\mu}\left(\frac{p^{\nu]}}{\sqrt{-p^2}}+\Lambda_0{}^{\nu]}\right)-\lambda\frac{\partial\mathcal M^2}{\partial S_{\mu\nu}}\right]\delta S_{\mu\nu}.
\end{align}
To maintain its orthonormality, the body-fixed tetrad $\Lambda_A{}^\mu$ must be varied according to $\Lambda_A{}^\mu\to(\delta^\mu_\nu+\delta\theta_\nu{}^\mu)\Lambda_A{}^\nu$, where $\delta\theta_{\mu\nu}$ is antisymmetric (giving an infinitesimal Lorentz transformation).  We find the linear variation
\begin{align}
\delta_\Lambda L&=\left[-\frac{1}{2}\frac{DS_{\mu\nu}}{ds}+S_{\rho[\mu}\Omega_{\nu]}{}^{\rho}+\Lambda_{0[\mu}S_{\nu]\rho}\chi^\rho \right]\delta\theta^{\mu\nu}
\nonumber\\*
&\quad+\frac{D}{ds}\left(\frac{S_{\mu\nu}\delta\theta^{\mu\nu}}{2}\right).
\end{align}
Finally, one can vary with respect to the worldline by letting $z$ move to a nearby point $\tilde z$ specified by a deviation vector $\xi^\mu$ at $z$, just as in Sec.~\ref{sec:shift}, while parallel-transporting $p$, $S$, and $\Lambda$ (and $\chi$) along.  The result for the linear variation, using (\ref{eq:ztildedot}) and (\ref{deltaOmega}), is
\begin{align}\label{dzL}
\delta_z L&=\left[-\frac{Dp_\mu}{ds}-\frac{1}{2}R_{\mu\nu\alpha\beta}\dot z^\nu S^{\alpha\beta}-\frac{\lambda}{2}\frac{\mathcal D\mathcal M^2}{\mathcal D z^\mu}\right]\xi^\mu
\nonumber\\
&\quad+\frac{D}{ds}\left(p_\mu\xi^\mu\right),
\end{align}
where the derivative $\dfrac{\mathcal D}{\mathcal D z^\mu}$ covariantly differentiates with respect to $z$ while parallel transporting $p$ and $S$. It is the ``horizontal covariant derivative'' of \cite{Dixon:1979} (where it is denoted $\nabla_{\mu*}$), and the covariant variation $\Delta$ of DeWitt \cite{DeWitt:2011,Steinhoff:2014} (see the following appendix) is $\Delta=\xi^\mu\dfrac{\mathcal D}{\mathcal D z^\mu}$.  If $\mathcal M^2$ depends on $z$ only through the metric and the Riemann tensor at $z$, then
\begin{equation}\label{hcd}
\frac{\mathcal D\mathcal M^2}{\mathcal D z^\mu}=\frac{\partial\mathcal M^2}{\partial R_{\nu\rho\alpha\beta}}\nabla_\mu R_{\nu\rho\alpha\beta}.
\end{equation}

Stationarity of the action requires that all four quantities in square brackets in (\ref{dpL}-\ref{dzL}) vanish.  Using the first three, one can eliminate $\Omega^{\mu\nu}$ and $\chi^\mu$ (needing only to solve for the projection of $S_{\mu\nu}\chi^\nu$ orthogonal to $p_\mu$), finding
\begin{equation}\label{Seomx}
\frac{DS^{\mu\nu}}{ds}=2p^{[\mu}\dot z^{\nu]}-\lambda\left(p^{[\mu}\frac{\partial\mathcal M^2}{\partial p_{\nu]}}+2S^{[\mu}{}_\alpha\frac{\partial\mathcal M^2}{\partial S_{\nu]\alpha}}\right),
\end{equation}
and then (\ref{dzL}) and (\ref{hcd}) yield
\begin{equation}\label{zeomx}
\frac{Dp_\mu}{ds}=-\frac{1}{2}R_{\mu\nu\alpha\beta}\dot z^\nu S^{\alpha\beta}-\frac{\lambda}{2}\frac{\partial\mathcal M^2}{\partial R_{\nu\rho\alpha\beta}}\nabla_\mu R_{\nu\rho\alpha\beta}.
\end{equation}
Contracting the $\delta_pL$ equation with $p_\mu$ yields
\begin{equation}\label{lambda_sol}
\lambda=\frac{p_\mu\dot z^\mu}{p^2}.
\end{equation}
We have assumed here that $\mathcal M^2$ depends on $p_\mu$ only through its direction $p_\mu/\sqrt{-p^2}$, implying that $p_\mu\partial\mathcal M^2/\partial p_\mu=0$, which is true for (\ref{masterM}).  If we do not assume this, it introduces $\mathcal O(S^2)$ corrections in (\ref{lambda_sol}).  The fact that $\mathcal M^2$ is a scalar implies that
\begin{equation}
p^{[\mu}\frac{\partial\mathcal M^2}{\partial p_{\nu]}}+2S^{[\mu}{}_\alpha\frac{\partial\mathcal M^2}{\partial S_{\nu]\alpha}}+4R^{[\mu}{}_{\rho\alpha\beta}\frac{\partial\mathcal M^2}{\partial R_{\nu]\rho\alpha\beta}}=0.
\end{equation}
Finally, with the identification
\begin{equation}\label{JMR2}
J^{\mu\nu\alpha\beta}=\frac{3p_\rho\dot z^\rho}{p^2}\frac{\partial\mathcal M^2}{\partial R_{\mu\nu\alpha\beta}},
\end{equation}
which is as in (\ref{JmcM}), we see that (\ref{Seomx}-\ref{JMR2}) yield the quadrupolar MPD equations (\ref{eq:MPDp}, \ref{eq:MPDS}).

One may also refer to the derivation in Ref.~\cite{Steinhoff:2014}, where a rather generic
action for spinning bodies is considered. In order to meet the requirements from \cite{Steinhoff:2014},
one must rewrite the Lagrange multipliers in the corotating frame as $\chi^\mu = \chi^A \Lambda_A{}^\mu$.
Then \cite{Steinhoff:2014} shows that (\ref{app:action}) leads to the MPD equations.

\section{Worldline shift from a covariant variation}\label{app:shift}
It is important to formulate a shift of the position in a manifestly covariant
manner, e.g., using bitensors as in Sec.~\ref{sec:shift}.
As a check, we rederive this shift using a manifestly covariant variation
symbol $\Delta$, defined in the previous section,
which was used in Refs.~\cite{DeWitt:2011,Steinhoff:2014} for obtaining
the equations of motion. It reads explicitly
\begin{equation}
\Delta=\delta+\Gamma^{\mu}{}_{\nu\alpha}\xi^{\alpha}G^{\nu}{}_{\mu},\label{covVar}
\end{equation}
where $G^{\nu}{}_{\mu}$ is a linear operator which rearranges spacetime
indices such that the covariant derivative $\nabla_{\alpha}$ can
we written in the abstract form 
$\nabla_{\alpha}:=\partial_{\alpha}+\Gamma^{\mu}{}_{\nu\alpha}G^{\nu}{}_{\mu}$. It holds
\begin{equation}
\delta z^\mu \equiv \xi^\mu ,
\end{equation}
which makes it manifest that $\delta z^\mu$ it a tangent vector and not a coordinate difference.
This is important when one considers finite worldline shifts.

We require the particle's properties to be parallel transported along
the geodesic connecting the two worldlines by setting
\begin{align}
\Delta \xi^\mu &=0,&\Delta p_{\mu}&=0, \\
\Delta S_{\mu\nu}&=0,&\Delta\Lambda_{A}{}^{\mu}&=0.\label{eq:shiftreq}
\end{align}
This implies that the component values of the worldline
quantities are actually transformed. However, since geometrically this
is just a parallel transport, we refrain from denoting this change
by a tilde on the indices in this section.
Notice that $\Delta \xi^\mu =0$ is just a restatement of the geodesic equation
for the worldline shift $\xi^\mu$.

Since the action is a scalar, the ordinary variation $\delta$ and the covariant
one $\Delta$ can be used interchangeably,
\begin{equation}
\mathcal{S}[z^{\mu}]
=\sum_{n}\frac{\delta^{n}\mathcal{S}}{n!}
=\sum_{n}\frac{\Delta^{n}\mathcal{S}}{n!}
\equiv e^{\Delta}\mathcal{S} ,
\end{equation}
where on the right hand side $\mathcal{S}$ is given by (\ref{eq:SGIaction}).
Only the functional form of $\mathcal{S}$ is important here, so that we
ignore the bars and tildes on the variables for now.
For the present paper the series stops at second order in $\xi^\mu$.
Useful formulas for the variation can be found in \cite{Steinhoff:2014},
for instance
\begin{align}
\Delta u^{\mu} &= \frac{D \xi^{\mu}}{d\sigma} , \\
[\Delta,D] &= R^{\mu}{}_{\nu\alpha\beta} \xi^{\alpha} dz^{\beta} G^{\nu}{}_{\mu} , \\
\Delta\Omega^{\mu\nu} &= R^{\mu\nu}{}_{\alpha\beta} u^{\alpha} \xi^{\beta} .
\end{align}
An application to the terms in (\ref{eq:SGIaction}) leads to
\begin{align}
\Delta(p_{\mu}\dot{z}^{\mu}) &= -\frac{Dp_{\mu}}{d s}\xi^{\mu}+\frac{D(p_{\mu}\xi^{\mu})}{d s} , \\
\Delta^{2}\left(p_{\mu}\dot{z}^{\mu}\right) &= - R^{\gamma}\,_{\mu\beta\alpha}\dot{z}^{\beta}p_{\gamma}\xi^{\alpha}\xi^{\mu} , \\
\Delta\left(\frac{1}{2}S_{\mu\nu}\Omega^{\mu\nu}\right) &= \frac{1}{2}S_{\mu\nu}R^{\mu\nu}\,_{\beta\alpha}\xi^{\alpha}\dot{z}^{\beta} , \\
\Delta\left(\frac{S^{\mu\nu}p_{\nu}}{p^2}\frac{Dp_{\mu}}{d\sigma}\right) &= R^{\gamma}\,_{\nu\beta\alpha}\dot{z}^{\beta}p_{\gamma}\frac{S^{\nu\mu}p_{\mu}}{p^2} \xi^{\alpha} .
\end{align}
After this shift was performed, one can again redefine the linear momentum
$p_\mu \rightarrow p_\mu + \delta p_\mu$. This reproduces (\ref{actiontransformation}), which completes the check
of the worldline shift. It should be stressed again that this shift represents
just a simultaneous transformation of all dynamical variables, understood as components,
at the level of the action. However, interpreted in geometric terms, the change of components
is just a parallel transport and the dynamical variables as geometric objects
remain invariant.

\section{Rotation coefficients for the curvature-aligned frame}\label{app:rot}

The components of the Ricci rotation coefficients for the curvature-aligned tetrad $e_a{}^\mu$ (\ref{CarterTetrad}), as in (\ref{omegaf}) and (\ref{omegath}) with $f\to e$, are given by
\begin{align}
\left({\omega}^{(e)}_{a*i}\right)&=\frac{\sqrt{\Delta}}{\Sigma^{3/2}}\left(\begin{tabular}{ccc}
 \ensuremath{\bar a_{1}} &  \ensuremath{\bar a_{2}} &  0 \\
0  &  0  &  \ensuremath{-\bar a_{1}\bar a_{2}/r}\\
0  &  0  &  \ensuremath{r}\\
\ensuremath{w_{3}r} &  \ensuremath{-r} &  0 
\end{tabular}\right),
\\
\left({\omega}^{(e)}_{a0i}\right)&=\frac{\sqrt{\Delta}}{\Sigma^{3/2}}\left(\begin{tabular}{ccc}
 \ensuremath{w_{0}r-\bar a_{2}^{2}/r} &  \ensuremath{\bar a_{1}\bar a_{2}/r} &  0 \\
0  &  0  &  \ensuremath{\bar a_{2}}\\
0  &  0  &  \ensuremath{\bar a_{1}}\\
\ensuremath{\bar a_{2}} &  \ensuremath{-\bar a_{1}} &  0 
\end{tabular}\right),
\end{align}
where we have defined 
\begin{align*}
w_{0} & =\frac{M}{r\Delta}\left(r^{2}-a^{2}\cos^{2}\theta\right),
\\
w_{3} & =\frac{\varpi^2}{r\sqrt{\Delta}}\cot\theta,
\\
\bar a_1&=a\cos\theta,
\\
\bar a_2&=-\frac{r}{\sqrt{\Delta}}a\sin\theta.
\end{align*}
We can express these in a 3-vector notation by forming the 4-vector
\begin{equation}\label{heth}
h^{(e)}_a=\omega^{(e)}_{a*i}S_i-\omega^{(e)}_{a0i} S_{0i},
\end{equation}
as in (\ref{hstar}), and defining the 3-vectors
\begin{align}\label{eq:S0r}
\vec{\bar a}=\left(\begin{tabular}{c}
 \ensuremath{\bar a_1}\\
\ensuremath{\bar a_2}\\
0 
\end{tabular}\right)
,\qquad\vec{r}=r\vec{n}=\left(\begin{tabular}{c}
 \ensuremath{r}\\
0 \\
0 
\end{tabular}\right),\qquad
\vec{w}=\left(\begin{tabular}{c}
0\\0\\
\ensuremath{w_3}
\end{tabular}\right),
\end{align}
along with $\vec S=(S_i)$ and $\vec S_0=(S_{0i})$, in the $e$-frame.  We find
\begin{align}
h_{0}^{(e)} & =\frac{\sqrt{\Delta}}{\Sigma^{3/2}}\left[-w_{0}\,\vec{r}\cdot\vec{S}_{0}+\vec{\bar a}\cdot\vec{S}-\frac{1}{r}(\vec{n}\times\vec{\bar a})\cdot(\vec{\bar a}\times\vec{S}_{0})\right],
\nonumber\\
\vec{h}^{(e)} & =\frac{\sqrt{\Delta}}{\Sigma^{3/2}}\bigg[-\vec{r}\times\vec{S}+\vec{r}\cdot\vec{S}\,\vec{w}+\vec{\bar a}\times\vec{S}_{0}\\
&\qquad\qquad \nonumber
+2(\vec{\bar a}\cdot\vec{n}\times\vec{S}_{0})\,\vec{n}
+\frac{1}{r}\vec{n}\cdot\vec{\bar a}\,(\vec{\bar a}\cdot\vec{n}\times\vec{S})\,\vec{n}\,\bigg],
\end{align}
from which, via (\ref{heth}), we can read off the components of the rotation coefficients.

\section{Hamiltonians through quadratic order in the spins to all PN orders}\label{app:allPN}

We present here the (untransformed) Hamiltonians through quadratic order in the test spin $S$ and the Kerr spin $S_\mathrm{Kerr}=Ma$, obtained as described at the beginning of Sec.~\ref{sec:expanded}, to all orders in the PN parameter $\epsilon\sim P^2/m^2\sim M/r$.  We retain here all factors of $m$.

At zeroth order in both spins, the point-particle Hamiltonian is given by
\begin{equation}\label{allpp}
H_\mathrm{p.p.}=\sqrt{w}\hat Q,
\end{equation}
where $w=1-2M/r$ as in (\ref{eq:w}), and
\begin{equation}
\hat Q=\sqrt{m^2+\vec P^{\,2}-\frac{2M}{r}(\vec n\cdot\vec P)^2},
\end{equation}
which is the $Q$ of (\ref{eq:PQ}) with $a\to0$.

The linear-in-spin Hamiltonians are
\begin{equation}\label{alla}
H_a=\frac{2M}{r^2}\vec n\cdot\vec P\times\vec a,
\end{equation}
and
\begin{equation}\label{allS}
H_S=\left(\frac{M}{r^2(\hat Q+m)}+\frac{\sqrt{w}-w}{r\hat Q}\right)\vec n\cdot\vec P\times\vec S.
\end{equation}

At quadratic order in the spins, we have
\begin{widetext}
\begin{align}\label{allaa}
H_{a^2}&=\frac{1}{\sqrt{w}\hat Qr^2}\bigg\{\vec a^{\,2}\left[-\frac{\vec P^{\,2}}{2}\left(1-\frac{8M^2}{r^2}\right)+\frac{(\vec n\cdot\vec P)^2}{2}\left(w(3-w)-\frac{8M^3}{r^3}\right)+\frac{2m^2M^2}{r^2}\right]
\\\nonumber
&\qquad\quad\qquad+(\vec n\cdot\vec a)^2w\frac{M}{r}\Big[m^2+2\vec P^{\,2}+w(\vec n\cdot\vec P)^2\Big]+\left[\frac{1}{2}(\vec P\cdot\vec a)^2-\vec n\cdot\vec P\;\vec n\cdot\vec a\;\vec P\cdot\vec a\right]\left(1-\frac{4M^2}{r^2}\right)\bigg\},
\end{align}
\begin{align}\label{allaS}
H_{aS}&=\frac{M}{r^3}\Bigg\{\vec a\cdot\vec S\left[2-3\sqrt{w}\left(1+\frac{\vec P^{\,2}-(1+w)(\vec n\cdot\vec P)^2}{\hat Q(\hat Q+m)}\right)\right]
\\*
&\qquad\quad+3\sqrt{w}\;\vec n\cdot\vec a\;\vec n\cdot\vec S\left[1+\frac{\vec P^{\,2}+\sqrt{w}(1-\sqrt{w})(\vec n\cdot\vec P)^2}{\hat Q(\hat Q+m)}\right]
\nonumber\\*\nonumber
&\qquad\quad+\frac{3\sqrt{w}}{\hat Q(\hat Q+m)}\left[\vec P\cdot\vec a\;\vec P\cdot\vec S-\vec n\cdot\vec P\;\vec n\cdot\vec a\;\vec P\cdot\vec S-(1+\sqrt{w})\,\vec n\cdot\vec P\;\vec P\cdot\vec a\;\vec n\cdot\vec S\right]\Bigg\},
\end{align}
and
\begin{equation}\label{allSS}
H_{S^2}=\vec S^{\,2}k_1(\vec z,\vec P)+(\vec n\cdot\vec S)^2k_2(\vec z,\vec P)+(\vec P\cdot\vec S)^2k_3(\vec z,\vec P)+\vec n\cdot\vec P\;\vec n\cdot\vec S\;\vec P\cdot\vec S\,k_4(\vec z,\vec P),
\end{equation}
where
\begin{align}\label{kcoeffs}
k_1&=\frac{M}{(\hat Q+m)r^3}+\frac{(1-\sqrt{w})^2r-2M}{2\hat Qr^3}\sqrt{w}+\left(\frac{M(2+\sqrt{w})}{2\hat Q(\hat Q+m)^2r^3}+\frac{(1-\sqrt{w})^2}{2\hat Q^3r^2}\sqrt{w}\right)\Big((\vec n\cdot\vec P)^2-\vec P^2\Big)
\\\nonumber
&\quad-(C-1)\frac{M\sqrt{w}}{2m^2\hat Qr^3}\Big(m^2-3(\vec n\cdot\vec P)^2+3\vec P^2\Big),
\\\nonumber
k_2&=M\frac{(\hat Q+3m)\sqrt{w}-\hat Q}{\hat Q(\hat Q+m)r^3}-\frac{(1-\sqrt{w})^2}{2\hat Qr^2}\sqrt{w}+\frac{M\sqrt{w}(w-2+\sqrt{w})}{\hat Q(\hat Q+m)^2r^3}(\vec n\cdot\vec P)^2+\left(\frac{M(1+2\sqrt{w})}{\hat Q(\hat Q+m)^2r^3}+\frac{(1-\sqrt{w})^2}{2\hat Q^3r^2}\sqrt{w}\right)\vec P^2
\\\nonumber
&\quad+(C-1)\frac{3M\sqrt{w}}{m^2r^3}\left(\frac{\hat Q^2+\vec P^2}{2\hat Q}+\frac{1-\sqrt{w}}{\hat Q+m}\sqrt{w}(\vec n\cdot\vec P)^2+\frac{(1-\sqrt{w})^2}{2\hat Q(\hat Q+m)^2}w(\vec n\cdot\vec P)^4\right),
\\\nonumber
k_3&=\frac{M}{\hat Q(\hat Q+m)^2r^3}+\frac{(1-\sqrt{w})^2}{2\hat Q^3r^2}\sqrt{w}+(C-1)\frac{3M\sqrt{w}}{2m^2\hat Qr^3}\left(1+w\frac{(\vec n\cdot\vec P)^2}{(\hat Q+m)^2}\right),
\\\nonumber
k_4&=-\frac{M(2+w)}{\hat Q(\hat Q+m)^2r^3}-\frac{(1-\sqrt{w})^2}{\hat Q^3r^2}\sqrt{w}-(C-1)\frac{3M\sqrt{w}}{m^2r^3}\left(\frac{1}{\hat Q}+\frac{\sqrt{w}}{\hat Q+m}+w(1-\sqrt{w})\frac{(\vec n\cdot\vec P)^2}{\hat Q(\hat Q+m)^2}\right).
\end{align}
\end{widetext}


%

\end{document}